# Laser Patterning of Superhydrophobic transparent glass surfaces with anti-fogging and anti-icing applications

Laura Montes-Montañez,[1] Fernando Nuñez-Galvez,[1] Melania Sanchez-Villa,[1] Luis A. Angurel,[2] Heli Koivuluoto,[3] German F. de la Fuente,[2] Víctor Trillaud,[4] Philippe Steyer,[4] Karine Masenelli-Varlot,[4] Francisco J. Palomares,[5] Ana Borrás,[1] Agustín R. González-Elipe,[1] Carmen López-Santos,[1,6] * Víctor Rico,[1]*

1 Nanotechnology on Surfaces and Plasma Lab, Institute of Materials Science of Seville (US-CSIC) Americo Vespucio 49, 41092 Seville (Spain)

2 Institute of Nanomaterials and Nanoscience of Aragon (UNIZAR-CSIC) María de Luna 3, 50018 Zaragoza (Spain)

3 Faculty of Engineering and Natural Sciences, Materials Sciences and Environmental Engineering, Tampere University, FI-33014 Tampere (Finland)

4 INSA Lyon, Universite Claude Bernard Lyon 1, CNRS, MATEIS, UMR5510, Villeurbanne, 69621, France

5 Institute of Materials Science of Madrid (CSIC), Sor Juana Inés de la Cruz 3, 28049-Madrid (Spain)

6 Departamento de Física Aplicada I, Escuela Politécnica Superior, Universidad de Sevilla, Virgen de África, 41011 Sevilla (Spain)

*mclopez13@us.es (C.L.S.); *victor@icmse.csic.es (V.R.)

**ABSTRACT:** This work addresses the fabrication of transparent glass surfaces with superior water-repellence (i.e., superhydrophobicity), and related functional properties such as omniphobicity, anti-fogging and anti-icing responses. Surfaces have been processed by means of mild femtosecond laser patterning combined with the grafting of fluorinated tethered molecules. Controlling the laser scan and ablation conditions permits the fabrication of under-design grooves with cross and lineal morphologies and separations between 10 and 500 μm. These patterns provide a high transparency and repellence to liquids and ice over large areas. The best performance is obtained for cross or parallel line patterned glass with microgrooves separated by 100 μm and 15 μm depth thanks to 5 laser scanning repetitions. These surfaces present a stable Cassie-Baxter wetting state and very low ice-adhesion strength while keeping up to 80 % of optical transmittance in the visible region. Anti-fouling tests, along with freezing and thawing cycles on these surfaces, have demonstrated a remarkable and durable self-cleaning response, even under environmentally stressful conditions. Water condensation experiments conducted under atmospheric conditions, as well as in an environmental electron scanning microscope, have revealed important issues related to the formation of supercooled water droplets. These experiments show that the patterned breakdown of surface properties effectively prevents extensive fogging and water accumulation. This feature is particularly crucial for low temperature applications where the preservation of transparency is essential.

## 1. Introduction

Glass is a universal material manufactured since the antique times that has been continuously growing in applications and new properties. For example, its modern use for window walls and display covers relies on a high transparency, chemical and environmental resistance, and the capacity of modern industry to prepare large area laminates. [1,2] Moreover, the functionalization of glass with the incorporation of protective, conductive, anti-reflective, low-emissive, or self-cleanable coatings has added specific functionalities of interest for industrial sectors such as energy, construction, transport, communication or electronics. [3] Unlike coated glasses, bare glass laminates are partially hydrophilic and prone to becoming covered with ice, fog, or fouling agents, events that affect the transparency and cleanness required for architectural glass, photovoltaic panels, or other optical systems working outdoors [4].

For non-transparent materials such as metals,[5] ceramics,[6] or even polymers,[7] different surface modification approaches have been developed to transform their surface into hydrophobic or superhydrophobic and to enhance their anti-stain,[8] water repellence,[9] or anti-icing properties.[10] These procedures usually involve the texturing and/or roughening of the material surfaces and the modification of their surface chemical composition.[11,12] However, these surface modifications inevitably produce a change in optical properties, [13,14] which, in the case of glass, can degrade its transparency due to light scattering, thus limiting its applicability in many sectors. Different studies have addressed the influence of carved motifs on the surface of glass on its light dispersion properties.[15,16]

Laser treatment has been used to generate the surface roughness and/or topographic texturing required to confer wetting and anti-icing functionalities to a variety of materials, [17,18] including glass. [19,20] The combination with grafting of tethered molecules has been also reported to stabilize hydrophobic or superhydrophobic responses. [18,21] These transformations have been generally discussed in terms of the Wenzel or Cassie-Baxter wetting regimes. [22] However, in laser-enabled superhydrophobicity, little or no attention has generally been paid to the possible modifications of optical properties, a key issue when dealing with glass. [23,24]

In this work, we present a procedure consisting of femtosecond laser patterning of microgrooves on the surface of borosilicate glass. The process relies on a single femtosecond laser, which is scanned several times over the glass surface. Such a feature facilitates scalability, surpassing alternative patterning methods combining the simultaneous effect of femtosecond-pulsed and continuous infrared lasers.[25] . Besides, a fine adjustment of microgroove characteristics is provided without damaging the glass plates. The patterned surfaces have been further grafted with tethered fluorinated molecules to bestow the surfaces with permanent superhydrophobicity, anti-icing and antifouling properties, as well as anti-fogging capacity (i.e., to control the water condensation behavior on the surface). A key requirement has been the preservation of high transparency in the visible spectral range. The research focuses on determining the relationship between the microgroove pattern characteristics and the optical properties, as well as the response towards wetting-icing (i.e., water in liquid, vapor, and ice forms) and biological liquid simulants.

Due to its importance for microfluidics [26] and water repellent surfaces, the study of water wetting properties with line microgrooves has deserved much attention, both from theoretical [27–34] and experimental [31,35–37] points of view. Results have been reported for aluminum, [31] ceramic, [36] or stainless steel [37] surfaces with lineal grooves, addressing questions such as the evolution of wetting and sliding angles of water droplets or the transition from Casie-Baxter to Wenzel wetting states. These analyses have determined the effect of the microgroove depth and width on hydrophobicity of the surfaces. Meanwhile, works intended for applications focusing on features such as the spreading easiness and their anisotropy of water droplets when sited on the microgrooves. Furthermore, grooved surfaces have been also used to enhance water condensation for liquid water collection purposes. [38,39] However, using grooved surfaces key issues of the present work such as the icing or the preservation of transparency are not specifically addressed.

The aim of this article is twofold. On one side, we present the parameters controlling the laser patterning and grafting procedures to provide multifunctionality to common glass plates. On the other side, we propose a rational criteria to categorize and understand the results obtained depending on grooves structure. Lineal and cross patterned surfaces have been manufactured and their wetting (to water and biological simulants), icing, and water condensation properties thoroughly characterized. Wettability to water and biological simulants for antifouling applications has been investigated through the determination of the apparent contact angle of liquid droplets and their sliding upon tilting the substrates.[40] The anti-icing performance has been assessed through freezing delay time and the ice-adhesion stress studies.[41,42] For water condensation analysis, we carried out experiments both under atmospheric conditions and in a scanning environmental electron microscope, in this latter case to monitor the formation and evolution of the first water moieties which form on the grooved surfaces. From this investigation, we have determined the pattern designs (lineal or cross) and groove separation that render the best trade-off between transparency and wetting and anti-icing responses, as required for outdoor applications. As a side result, we also show the compatibility of the patterned procedure with the hosting of lubricant liquids to harness superior transparency and outstanding water slipperiness.

## 2. Materials and methods

*Materials and samples preparation*. Borofloat®33 (VIDRASA S.A.), a floated borosilicate flat glass, was used as transparent substrate. Borosilicate glass composition includes silicon dioxide ($SiO_2$), boric oxide ($B_2O_3$), and aluminum oxide ($Al_2O_3$) in a range from about 60 % to 74 %, 9 % to 25 %, and 7 % to 17 % weight, respectively. The presence of $B_2O_3$ provides a higher chemical stability to borosilicate glass as compared to soda-lime glass, [43] and improves its mechanical flexural strength. It also exhibits lower thermal expansion values, making it less susceptible to mechanical failure due to sudden thermal shocks. This is an additional reason for choosing borosilicate glass for laser processing and outdoor applications.[44]

Schematic 1 shows the experimental procedure scheme utilized in this work. Borosilicate plates with dimensions of 10 cm x 5 cm were cleaned using detergent water solutions, isopropanol, and a $N_2$ dry flow. Laser patterns were generated with a femtosecond (fs) laser (Carbide model, Light Conversion, Lithuania), using its third harmonic (UV range) at 343 nm. The laser beam exhibited a spatially Gaussian energy profile with an elliptical shape, where $1/e^2$ beam dimensions were $2\,a = 60$ μm (long semi-axis) and $2\,b = 36$ μm (short semi-axis) (Schematic 1 II). This laser offers the possibility of using a Pulse Peak Divider (PPD) to reduce the effective frequency of the laser treatment. The following working conditions have been established after a preliminary optimization step: 6.07 W laser power, 200 kHz oscillator frequency, 10 kHz effective frequency, 25 mm/s laser scan rate, and 238 fs pulse duration. Similar conditions were used in a previous work where laser parameters were optimized to generate superhydrophobicity on PU-based surfaces, similar to those used in wind generator blades.[45] The corresponding energy per pulse resulted in 30.4 μJ/pulse. Two hatching configurations were tested leading to cross and line patterns with distances between scanning lines ranging between 10 μm and 500 μm. Considering that the laser scanning speed is the same in both directions, when the laser moves parallel to the $2b$ semiaxis, the maximum fluence at the centre of the line was 53.6 J/cm$^2$. Meanwhile, when the laser moves parallel to the $2a$ semi-axis, the overlapping between consecutive pulses is lower and the maximum fluence is reduced to 32.1 J/cm$^2$. Depending on the experiments, the laser scanning protocol was repeated between 1 to 5 times. The width of the laser track depends on the scanning direction, parallel either to the short or the long semiaxis (see Schematic 1 II). For 5 laser scans, the groove widths were 37 μm and 25 μm, respectively.

Laser treated borosilicate surfaces were functionalized by grafting of 1H,1H,2H,2H-Perfluoro-octyltriethoxysilane 98 % (PFOTES) molecules according to the procedure described in reference [46] (Schematic 1 III). Borosilicate samples used as reference have been designated as *glass ref*, and the laser treated samples have been identified according to the following labelling: laser cross and line patterned surfaces as *C-distance* and *L-distance*, respectively, where a "distance" value expressed in microns corresponds to the distance between the centres of two contiguous laser scanned grooves; labels

*F-glass* (for flat glass), *F-C-distance* and *F-L-distance*, refer to samples after the molecular fluorination grafting. The actual number of scans has been added to the previous labelling as (*number*). Finally, *the K-C/L-distance* label designates a patterned surface, which was infiltrated with a fluorinated lubricant: Krytox[R] 100 (DuPont), with a refractive index around 1.296-1.301. Infiltration was done according to the procedure described in [47].

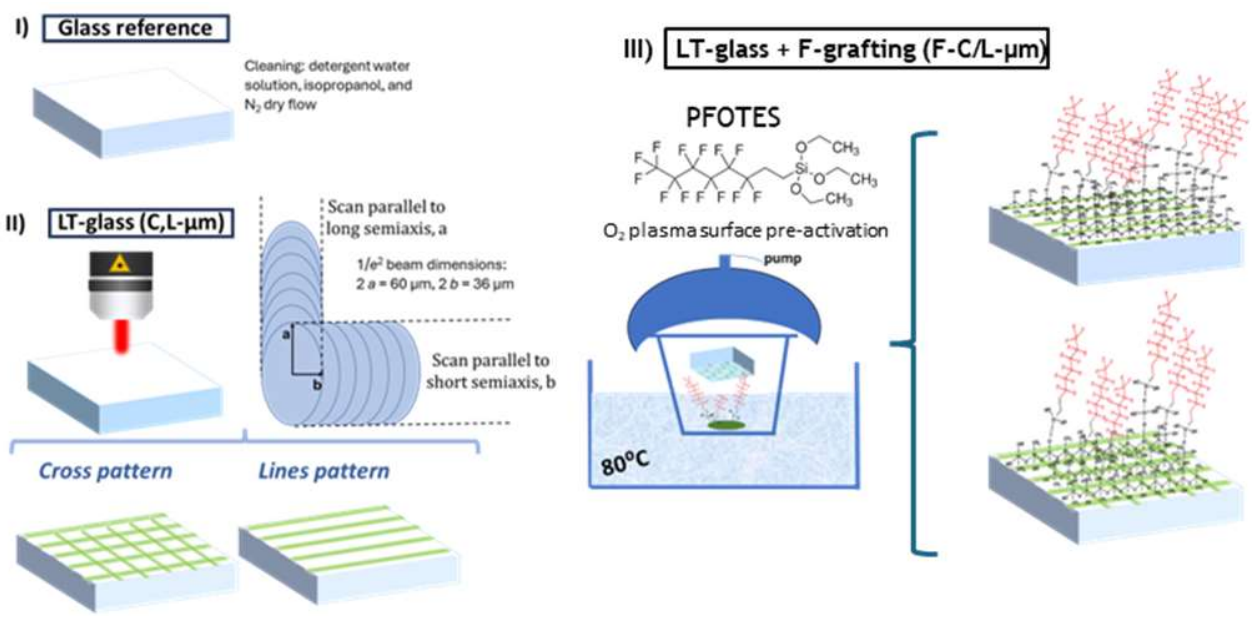

**Schematic 1**. **Experimental steps to produce laser treated and fluorinated borosilicate glass**. I) Cleaning of the as received glass; II) laser treatment to generate rectangular or parallel line patterns. A scheme of the laser spot along the two scanning directions highlights the long and short elliptical semi-axes of the laser spot; III) grafting with perfluorinated molecules of laser patterned surfaces.

*Surface characterization and functional properties*. Surface morphology of reference and laser-treated samples was examined by scanning electron microscopy (SEM, Hitachi S4800) at 2 kV, and confocal microscopy (Sensofar PLµ2300). Arithmetical ($S_a$) and mean square root ($S_q$) roughness parameters, have been determined from confocal optical images: the first one represents the arithmetic mean of the topography heights relative to a baseline and provides a general measure of roughness; the second one is the square root of the mean of the topographic tips and hillocks relative to the baseline. Width and depths of the patterned surfaces were estimated as the average over 3 different measurements. ImageJ was used for the calculation of the percentage of affected surface area over 2D images directly affected by the laser treatment.

The optical properties of the processed glass samples were determined in transmission mode in the UV-vis-nIR range between 200 and 900 nm using a Perkin Elmer Lambda 750 spectrometer equipped with a 60 mm diameter integration sphere.

Surface chemical analysis was carried out by X-ray photoemission spectroscopy (XPS) in a SPECS spectrometer (DLSEGD-Phoibos-Hsa3500) using nonmonochromatic Al Kα radiation. Data were recorded in constant pass energy mode at 50 eV for the general survey spectra and 30 eV for high-resolution spectra. Calibration in binding energy (BE) was done at the carbon functional C–H and C–C bonding groups appearing at 284.5 eV in the C1s region. XPS experiments with lateral resolution were performed in a UHV chamber with a base pressure of $10^{-10}$ mbar equipped with a hemispherical electron energy spectrometer (Phoibos 150, SPECS Surface Nano Analysis GmbH, Germany) and a 2D delay-line detector (Surface Concept GmbH, Germany), using an X-ray source of Al-Kα (1486.6 eV). XPS spectra were acquired at normal emission take-off angle, using an energy step of 0.50 and 0.10 eV and a pass-energy of 40 and 20 eV for survey spectra and detailed core level regions, respectively. Chemical analysis was performed recording the intensity of O 1s, Si 2p, F 1s, and C 1s peaks using pass energy of 100 eV.[48] The effective pixel size was 30 µm x 6.25 µm, determined by the selected slit, magnification factor, and detector active area selection. XPS analysis with lateral resolution was carried out moving the sample along the X-direction in 20 µm steps over 300 µm. The spectra were analyzed with the CasaXPS program (Casa Software Ltd., Cheshire, UK [49]) using a Shirley method for background subtraction and data processing.

Wetting properties were characterized by the sessile droplet method in an OCA 20 DataPhysics goniometer, depositing different volumes (2 µl for apparent static contact angle (CA) and 15 µl for rolling-off angle (RoA)). Bi-distilled water and different organic liquids as fouling simulants, including humic acid, sodium alginate and bovine serum (Sigma Aldrich) were used for this analysis. The reported data are an average of at least 10 experimental measurements per surface condition. Freezing Delay Time (FDT) measurements were performed in the OCA 20 DataPhysics controlled environmental chamber. The surface temperature was stabilized with a thermoelectric module and was tracked with a thermocouple placed on the surface of the samples at 1 mm from the water droplet. A flow of dry nitrogen (~10 sccm) was cooled at 0 $^0$C before passing through the chamber. The 2 µl water droplets were deposited on the surface at 25 °C. Then, the system was allowed to cool down to the desired freezing temperature. The time required for icing was monitored when the sample temperature had reached the nominal temperature of the experiment. Icing temperature was set to -10 $^0$C and it was reached applying a variation ramp of -1 °C/s until 0 $^0$C and -0.1 °C/s in the subzero temperature range.

Water condensation experiments were carried out in an OCA 25 DataPhysics environmental chamber provided with a temperature control module and a humidity generator. Experiments were performed with the samples in vertical configuration. In-situ water condensation at (sub)micron scale was also characterized in Environmental Scanning Electron Microscopy (ESEM) (Quanta FEG250 system from FEI and Thermofisher QuattroS) by placing the sample in a horizontal position on top of a Peltier stage, while controlling the environmental water vapor pressure. The temperature was stabilized at 2 °C and the water condensation was monitored as a function of water vapor pressure. Walls of the microscope chamber were not cooled, only the sample on the Peltier stage. An estimate of the condensed volume was deduced from the treatment of scaled images, accounting for the surface covered by the area of almost spherical drops.

Ice adhesion tests were carried out at -13 $^0$C using the pull-off method within a climate chamber acting as universal icing material testing machine (Signeblock a9fe07d986f95ae20575d9bf1ad2665ca4b8074e7b65df6dd8d4bbe 33e9b8d51). A Teflon cylinder with an internal diameter of 9.86 ± 0.12 mm and an area of 76.4 ± 1.9 mm$^2$ was filled with water up to approximately 13 mm. It was then brought into contact with the cooled samples within low humidity environmental conditions. After icing, the force applied perpendicular to the sample (tensile mode) in order to detach the ice block from the surface was determined with an IMADA ZTA-200n/20N dynamometer linked to a motorized IMADA MHZ-500N-FA linear stage. The adhesion strength was obtained by normalizing this force with respect to the area of the ice cylinder and expressed in units of MPa ($N/m^2$). Preliminary ice adhesion tests in pushing configuration were done at TAU in Ice Laboratory, where ice was accreted using an icing wind tunnel. Icing parameters were as the temperature of -10 ºC, a droplet size of ~30 μm and a wind speed of 25 m/s. Ice thickness was targeted as 10 mm and a mixed glaze ice was used in the experiments. Ice adhesion was measured with a pushing type ice adhesion tester, which operates with a force gauge (Sauter FL 500, Kern & Sohn GmbH, Balingen, Germany) and manual pushing holder. Tests were done in cold conditions (-10 ºC), where ice-accreted samples were kept ~16 h before testing to ensure full freezing.

## 3. Results and discussion

### 3.1 Surface morphology and chemical composition by laser patterning

Regular replicas of surface grooves with reproducible size and depth at the microscale (i.e., microgrooves) were obtained over large surface areas of borosilicate glass (10 cm x 5 cm). Laser overlapping and scanning direction have been the main process variables optimized for the generation of groove patterns. The distance between the centre of microgrooves was taken as a ruling parameter to correlate with the wetting response of laser-treated glass.[50,51] **Figure 1** shows selected confocal images for the obtained designs: parallel lines and rectangular micropillars, the latter resulting from laser cross-scans in normal directions. To ensure the reproducibility of the microgroove depth and width, the laser scanning process was repeated up to five times. Figure 1 panels a)-d) correspond to a selection of images of the *L-50/100(5)* and *C-50/100(5)* samples (see Experimental Section). Differences in the microgroove depths were obtained as a function of the number of scans, as exemplified in Figure 1 e)-h) for samples *C-100*.

In the case of *L* samples, the laser scan direction was parallel to the short laser beam semiaxis, *2b*. Regular microgrooves with 13 μm depth and 37 μm width at the surface were obtained after 5 scans (Figure 1 a) and b)). For *C* samples, microgroove features differed for the two scan directions. When the scanning proceeded parallel to the long spot semiaxis *2a*, the accumulated laser fluence was higher. After five scans in this direction, the obtained microgrooves exhibited a shorter width, i.e., 26 μm (note that the laser spot diameter is around 25 μm) and depths around 12-15 μm. Conversely, when the scan direction was normal to the long spot semiaxis, the accumulated laser fluence decreased accordingly and, after five repeated scans, grooves similar to those developed in the L samples were obtained (Figure 1 c) and d)). This evidences that, although rectangular motifs preserve their dimensions (~74 μm x 63 μm for 100 μm of laser groove separation), the anisotropy between laser scanning at normal directions becomes more significant for 5 scan repetitions, especially for the magnitude of scratched depths of grooves. Thus, the resulting microgrooves along the direction parallel to the long laser beam semiaxis were wider, whereas the microgrooves generated in the normal direction were narrower and displayed a similar or even slightly increased depth. As a result of the elliptical shape of the laser beam, microgroove depth and width were also dependent on the number of scans, from 1 to 5 repetitions (see **Figure S1** and **Table S1**), with width values varying from 24(20) μm to 37(26) μm and depth values from 2(3) μm to 10(15) μm, for the laser scanning along the short(long) semiaxis for an increased number of repetitions. Furthermore, the greater the depth caused by laser overlapping, the greater the narrowness developed at half the height of the generated groove (Figure S1 and Table S1). The non-linearity in the variation of the groove dimensions with the number of scans has been attributed to a change in the laser absorption capacity by previously treated surfaces.[45]

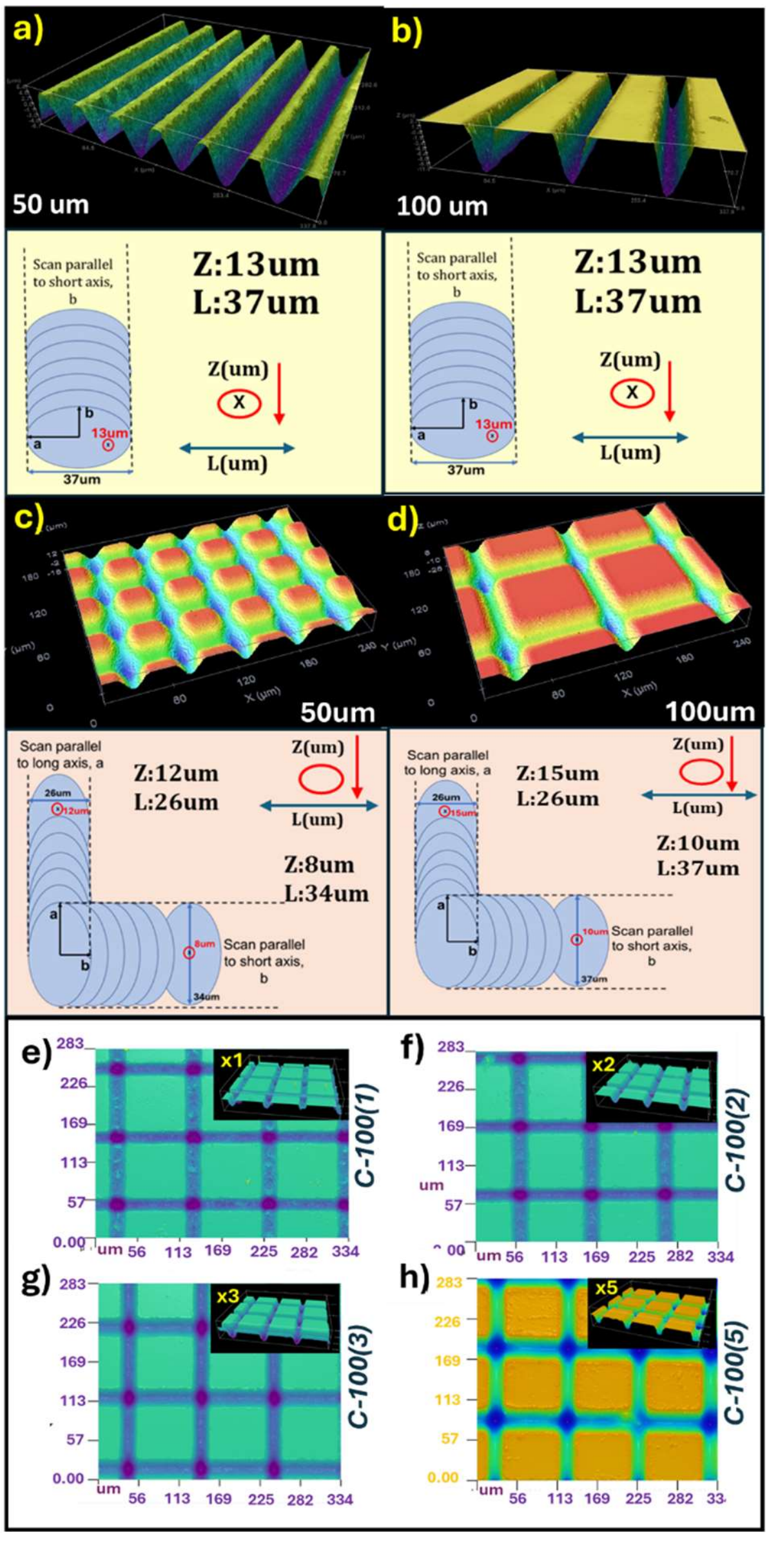

**Figure 1. 3D maps for representative laser-treated borosilicate glass.** Confocal microscopy maps for (a) *L-50/100* and (c) *C-50/100* samples after 5 laser scan repetitions; (b) and (d) show the corresponding details for the laser scans in (a) and (c). Depth and width are indicated for each laser direction (*a* and *b* long and short semi-axles); e)-h) 2D maps (arbitrary colour scale bar) for sample *C-100* for different numbers of laser scans repetitions from 1 to 5.

It is worth noting that after 5 scans, we encountered significant changes in roughness. The particular rough morphology of the microgroove valleys can be associated with material melting/solidification phenomena which, in addition, hinder the progression of the groove depth.[25] To account for the effect of the laser treatment on the surface properties, we have estimated the percentage of laser affected surface, i.e., the relative surface area occupied by the microgrooves (see **Table 1**).

**Table 1**. laser affected surface area percentages normalized to a 1 $cm^2$ for laser patterned glass surfaces obtained from digital imaging analysis of 3D confocal maps.

| Lines (*L*-) samples | Laser-treated surface % | Cross (*C*-) samples | Laser-treated surface % |
|---|---|---|---|
| L-10(5) | *100%* | C-10 | *100%* |
| L-25(5) | *100%* | C-25 | *100%* |
| L-50(5) | *50%* | C-50 | *87%* |
| L-100(5) | *25%* | C-100 | *53%* |
| L-200(5) | *13%* | C-200 | *29%* |
| L-300(5) | *9%* | C-300 | *19%* |
| L-500(5) | *5%* | C-500 | *12%* |

This parameter has been demonstrated as more relevant to correlate optical and wetting properties than the average surface roughness.[45] Looking at Table 1, it is remarkable that for 10 μm and 25 μm samples, no untreated area was observed to remain on the glass surface, i.e., the microgrooves overlap affecting the entire area of the samples. The affected surface area percentages decreased drastically with the microgroove separation reaching minima for 500 μm distances both for *C*- and, particularly *L*- samples. A general result stemming from Table 1 is that for the same distance between microgrooves, the percentage of laser affected surface area in *C*- samples is always higher than in *L*- samples.

To support these findings, the SEM images in **Figure 2** show that, due to the superposition of microgrooves, sample *L-10(5)* presented no pattern, but a continuous rough surface characterized by a microfeatures morphology (Figure 2 a)). In *L* samples, only for a distance between lines greater than the size of the laser spot (25 μm), a pattern of defined microgrooves begins to emerge on the surface. High magnification SEM micrographs revealed the formation of elongated nanostructured features of around 1 μm long, and 300 nm width at the bottom region as well as the walls of the grooves, independently of the distance between them(see also **Figure S2** at Supporting Information).

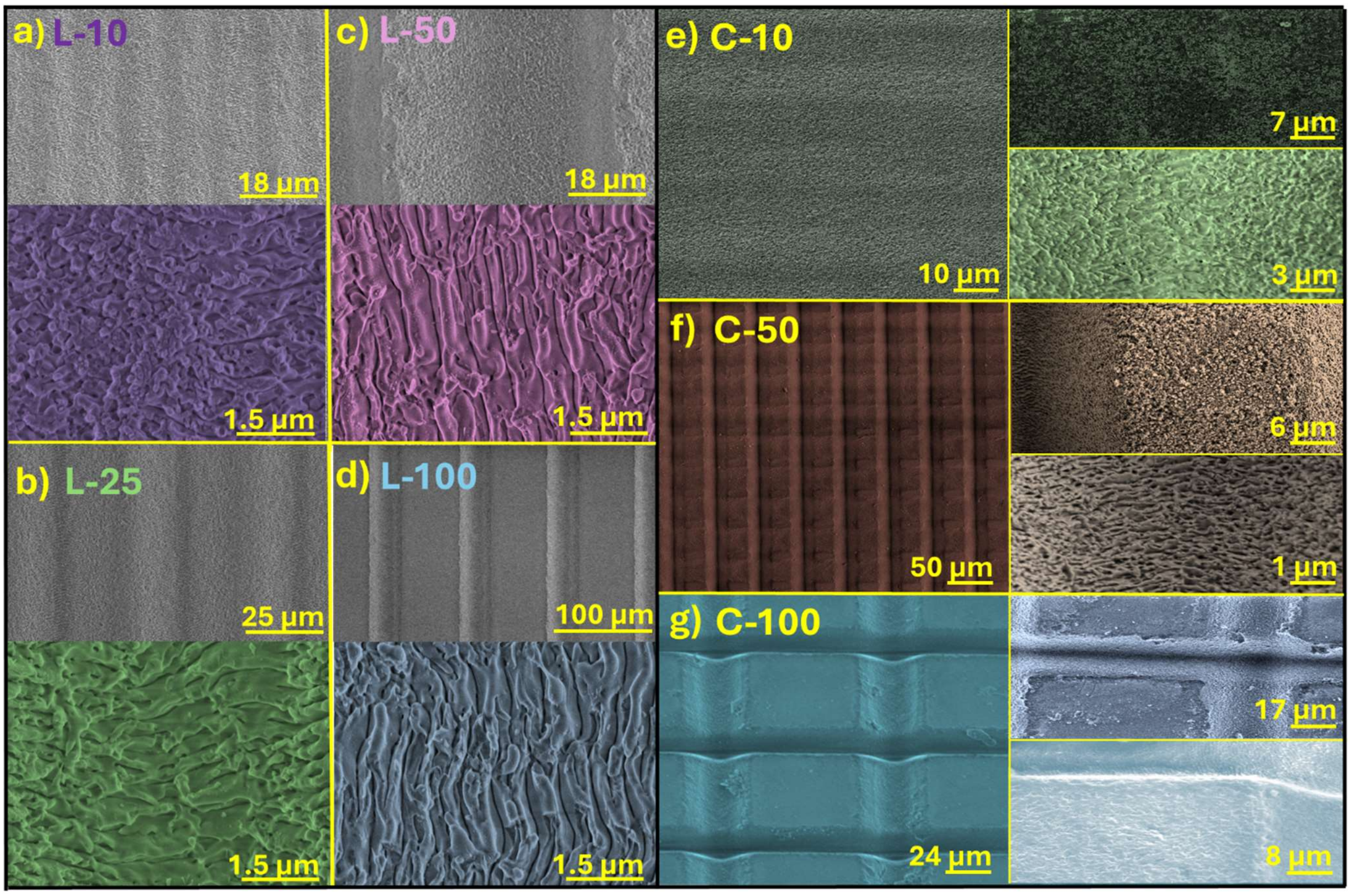

**Figure 2. SEM characterization of laser patterned surfaces.** SEM micrographs at different zoom scales for patterned samples: a) *L-10(5)*, b) *L-25(5)*, c) *L-50(5)*, d) *L-100(5)*, and e) *C-10(5)*, f) *C-50(5)*, and g) *C-100(5)*. Laser treatments were carried out applying five laser repetitions to draw the different motifs.

These observations in surface morphology are also reflected in surface roughness, expressed in terms of the $S_a$ and $S_q$ parameters (**Table S2**), which increased with the number of laser scan repetitions and reached maximum values for *L-100(5)* and *C-100(5)* samples (~ 4 μm and 3 μm for *Sa* and *Sq*, respectively). Then, for larger distances between grooves, constant roughness values around 1 μm ($S_a$) and ~2.3 µm ($S_q$) were measured. These results can be correlated with the almost constant percentage of laser affected surface area in these samples (c.f. Table 1). *C-* samples illustrate the previous topographic tendencies. For example, in sample *C-10(5),* no pattern was developed because of the superposition of laser lines, while sample *C-50(5)* depicted well-defined rectangular zones of untreated surface. Nanostructuring also developed along the walls and inside the microgrooves (see a more detailed analysis in **Figure S3** of the Supporting Information). Therefore, also in this case, a surface with features at different scales defines a hierarchical topography resulting from the combination of micro-(laser patterned microgrooves) and nano-(inside the microgrooves) structures.

Surface composition determined by XPS for laser patterned samples revealed a small increase in the carbon content for laser patterned samples, while the O/Si ratio remained virtually constant with values around 2, typical of $SiO_2$ (see **Figure S4** and **Table S3** in Supporting Information). Meanwhile, after molecular grafting, data also showed that the surface functionalization led to the incorporation onto the glass surface of a considerable amount of tethered PFOTES molecules, rendering fluorine concentrations of 21.4 and 27.8 atomic % on the untreated and laser-treated surfaces, respectively. This fluorine content appeared in the form of $-CF_2$ and $-CF_3$ functional groups[52] (Figure S4 b)). Further XPS analysis with lateral resolution evidenced a rather uniform distribution of fluorine along the complete patterned surfaces, suggesting no preferential grafting of PFOTES molecules on the laser-induced microgrooves.

### 3.2 Optical properties of laser patterned glass surfaces

An overview of the UV-visible transmittance spectra of patterned glass samples is gathered in **Figure 3**. A first evidence is that the surface grafting process did not appreciably affect the optical properties neither of the *L* nor the *C* samples (see also **Figure S5** in the Supporting Information). A general assessment from the set of spectra of samples *F-L* (Figure 3 a)), indicates that they maintain a high degree of transparency in the 550-800 nm visible range, with just a slightly reduction in the visible region at wavelengths smaller than 500-550 nm, but experience a significant transmission decrease in the UV region. This decrease reached maximum values of 40% and 20% for samples *F-C-100(5) and F-C-50(5),* respectively. Samples *F-C-200/300/500(5),* characterized by a percentage of laser affected surface fraction smaller than 30 % (c.f., Table 1) presented a transmission higher than 80% in the visible region. Samples *F-C-25(5)* and *F-C-10(5)* transmitted well the light for wavelengths higher than 550 nm but depicted a pronounced decrease in the UV region. Loss of transmittance can be related to light scattering effects, which are more important at low wavelengths. For patterned surfaces, it is known that such light scattering effects depend on feature shape, size, and distribution.[53,54,55] A similar behaviour identified in laser processed glass has been related to Mie scattering in the roughness features.[56] In our case, the magnitude of the UV transmission decrease can be correlated with the percentage of laser surface affected area. This empirical correlation fails for samples *F-L-25(5),* where no well-defined microgrooves are formed. The high loss of transmission found for this sample might be due to its high average roughness, one of the highest of the whole series of studied samples (see Table S2).

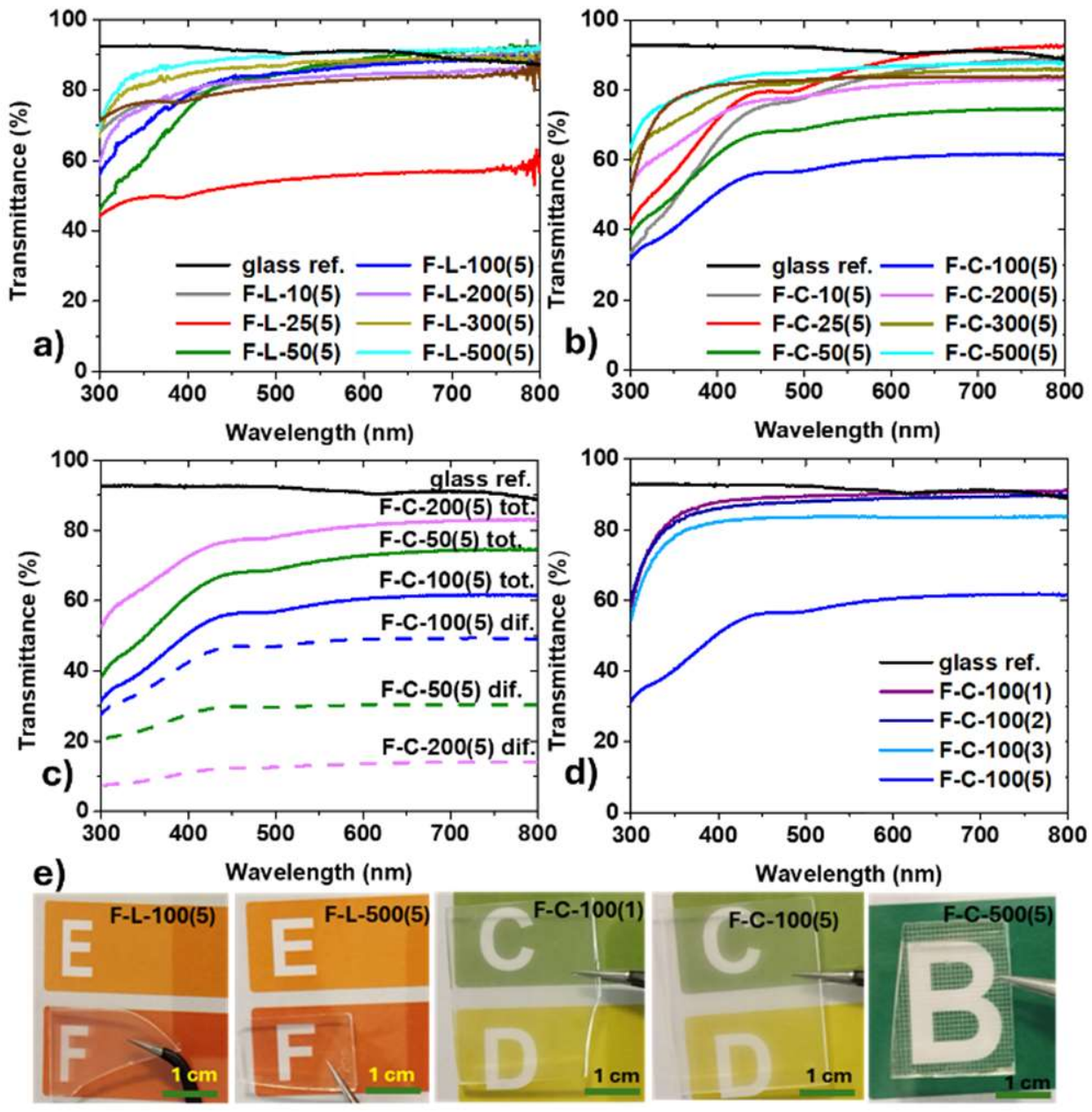

**Figure 3. UV-visible transmission spectra of reference, and patterned fluorinated glass samples**. Total transmittance of *F-L* a) and *F-C* b) samples as a function of the microgroove separation; c) examples of the total (tot.) and diffuse (dif.) transmittances for *F-C-* surfaces as a function of the separation between microgrooves; d) total transmittance spectra of samples *F-C-100* depending on the number of laser scans; e) photographs to directly assess the transparency of the patterned glass samples.

Figure 3 b) reveals a similar optical transmission behaviour for samples *F-L* and *F-C*, albeit an additional transmission decrease in the UV and near UV visible regions in the latter case that agrees with the smaller fraction of non-treated surface areas in the *C* samples. For selected *C* samples, it can be also seen in Figure 3 c) that the decrease in ballistic light transmission is directly related with a non-negligible diffuse transmittance component, which is particularly high for sample *F-C-100(5).*

The transmission spectra in Figures 3 a) and b) correspond to patterns drawn after five times laser scanning. Further analysis revealed that the optical transmission increased for patterns manufactured with less scans. Figure 3 d) presents this effect for the series of samples *F-C-100*. Remarkably, the recorded transmission spectra for samples *F-C-100(2)* and, particularly *F-C-100(3),* approach that of the reference sample for the whole range of visible wavelengths. This tendency suggests that the decrease in optical transmission with the number of laser scans is due to the progressive increase in depth and nanostructuring inside the microgroove (Figure 2). It is also important to stress that even for sample *F-C-100(5),* which presented the smallest light transmittance in the visible, photographs included in Figure 3 e) demonstrate sufficient transparency for applications as transmission windows or displays.

A known procedure to decrease light scattering effects in rough surfaces is through the addition of liquids with a refraction index similar to that of glass.[57] To demonstrate the feasibility of this strategy with the laser treated glass samples, a fluorinated lubricant, Krytox (refractive index 1.296-1.301), was infiltrated in two laser patterned samples (*F-L-25(5)* and *F-C-100(5))* leading to the formation of slippery liquid-infused porous surfaces (SLIPS). The optical characterization of these slippery surfaces revealed an enhancement in transmission up to values of ca. 80% for the UV and visible spectral regions (see **Figure S6** in the Supporting Information), demonstrating that liquid infusion is a suitable procedure to enhance optical transmission in these systems.

### 3.3 Wetting study of transparent patterned glass surfaces

Pristine glass surfaces presented apparent WCAs lower than 60° for droplets of 2 µl. Meanwhile, as-prepared samples *L-* and *C-* were superhydrophilic as expected for rough surfaces under the premises of the Wenzel model.[58] This superhydrophilic state lasted for more than 3 months for samples stored under room conditions. To permanently convert the patterned samples into hydrophobic without affecting their optical properties,[59,60] we proceed to graft perfluorinated molecules over the borosilicate surface. This procedure has been previously used for the fabrication of optimal self-cleaning, anti-fouling and anti-icing alumina[61] and stainless-steel hierarchical surfaces[5] with a minimum content of perfluorinated molecules anchored to the surface to reach a stable repellent behaviour. Herein, it was found that grafting PFOTES molecules onto a flat glass surface induced a net and permanent increase in WCA from 57° to 105° (c.f., **Figure 4**). Similarly, the *F-L* and *F-C* grafted samples became highly hydrophobic or even superhydrophobic (see Figure 4 and **Tables S4** and **S5** in the Supporting Information). The achieved hydrophobic or superhydrophobic states agree with the behaviour expected for hierarchical rough surfaces with a high intrinsic hydrophobicity as provided by the grafted molecules. A similar effect was reported by Li et al. [36] and Douyan et al.[56] for, respectively, laser patterned ceramics or glass grafted with fluorine compounds. Looking in detail to the behaviour of patterned and grafted samples, it appears that all samples *F-L(5)* were hydrophobic with apparent static WCAs higher than 100°, approaching 150° for sample *F-L-10(5)* and a superhydrophobic state for samples *F-L-25(5)* (c.f., Figure 4 a)). As discussed in section 3.1, in these two samples the microgrooves overlap in such a way that their surface does not preserve flat zones between microgrooves. Interestingly, Figure 4 a) shows that the apparent WCAs determined for *F-L-50(5)* to *-500(5)* samples progressively decrease with the percentage of laser affected surface area (c.f., Table 1). This behaviour is congruent with current wetting models in heterogeneous surfaces consisting of flat parts behaving as hydrophobic and nanostructured zones (i.e., the microgroove walls and valleys) as superhydrophobic, i.e.:

$$cos\theta^{CB} = \sum_i f_i cos\theta_i^Y \qquad (1)$$

where $\Theta^{CB}$ is the apparent contact angle under the premises of a Cassie-Baxter model and $\theta^Y$ is the contact angle of each of the different zones *i* of the surface (Young state). [29]

It is also noteworthy in the experiments with the *F-L* samples that the small (i.e., 2 µl) water droplets used to determine the apparent WCAs depicted a circular shape with no anisotropy along the microgroove direction. The isotropic character was further confirmed by the fact that no sliding of large 15 µl water droplets was found on the majority of *F-L* samples with the exception of *F-L-500(5)* surface where water droplets readily slid along the microgrooves (RoA of 23°, see Supporting Information, Table S4). With respect to other works reporting that linear grooves induce anisotropy and directional sliding of deposited water droplets, [31,35–37] the circular shape of droplets and the lack of sliding must be attributed to the relatively smaller depth of microgrooves [45] in our samples, a requirement dictated by the criterium of transparency, as discussed in the previous section.

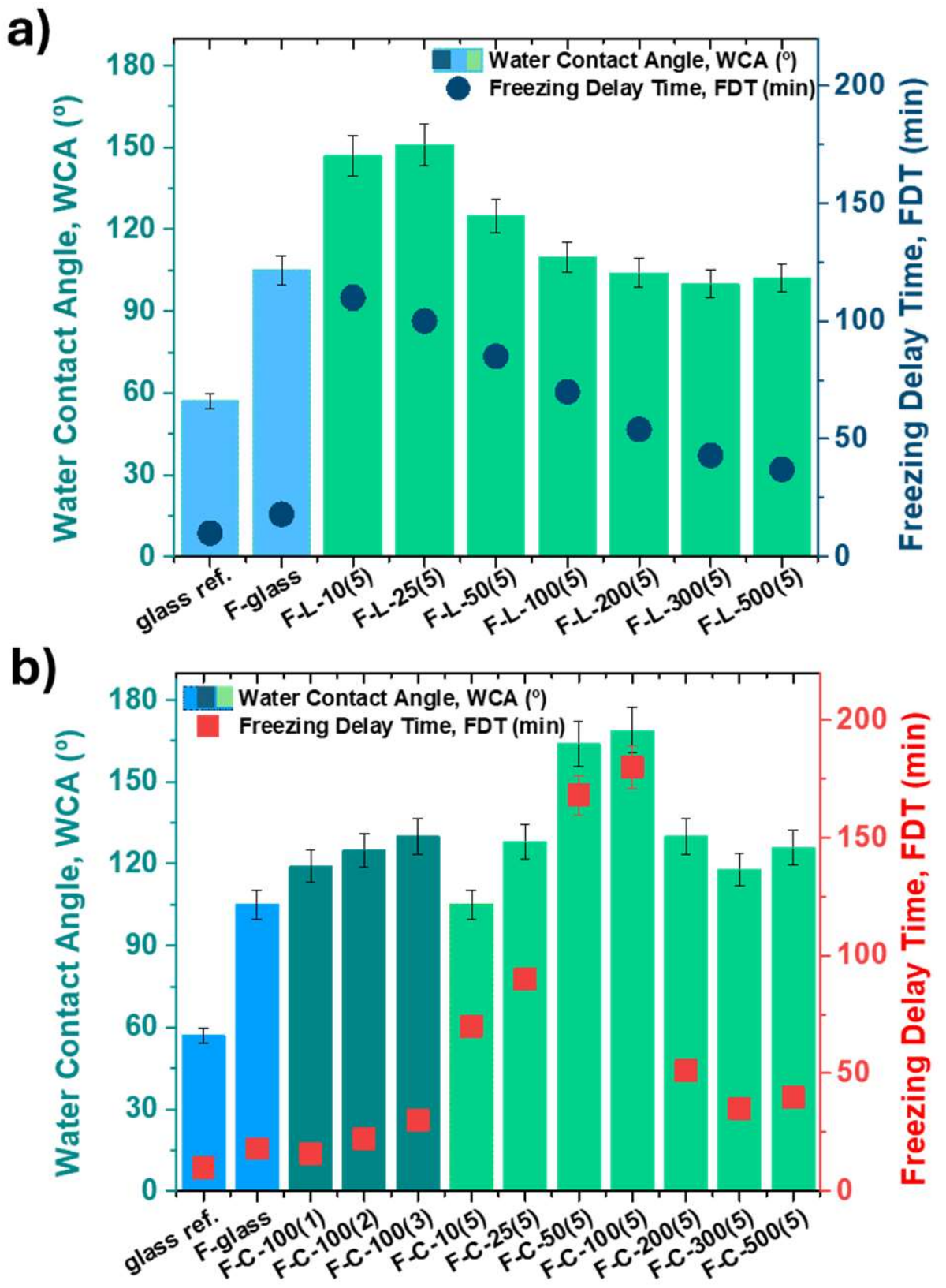

**Figure 4**. **Wettability and freezing delay times of laser-treated surfaces.** Apparent water contact angles (columns) and freezing delay times (spots) for a) *F-L-* and b) *F-C-* samples, including the effect of the number of laser scans. Data for an untreated flat glass used as reference are included for comparison.

The apparent WCAs of *F-C* samples are reported in Figure 4 b). The variation of this parameter with the number of laser scans from 1 to 5 for samples *F-C-100* is also included. All *F-C(5)* samples were hydrophobic (generally with a higher WCA than their equivalent *F-L* samples). This renders a state close to superhydrophobicity for samples *F-C-50(5)* and *F-C-100(5)*. In these two samples, the percentage of untreated laser surface area is smaller than in the equivalent *F-L* samples (c.f., Table 1), while the microgroove depth, width, and the roughness inside the microgroove is different for each direction, features that provide optimum conditions to convert the glass surface into superhydrophobic[62] under a Cassie-Baxter regime.[45] It is also noteworthy that for these two samples the rolling-off angle was as low as 2° (see Table S5 in the Supporting Information), ensuring an effective sliding of water. A similar behaviour was reported for fluorinated rectangular microgroove patterns by Zhong et al.[39], who attributed the enhanced sliding to a chemistry homogenization of the spatial wetting response. Figures 4 a) and b) also show that the apparent WCAs of the other samples of the *F-L* and *F-C* series progressively decrease with the distance between microgrooves, a behaviour that can be accounted for by the premises of eq. (1).[63] A rational hypothesis for the observed differences is schematically described in **Figure 5**, comparing the wetting of sample *F-C-100(3)*, *F-C-100(5)* and any *F-C-* samples with very separated grooves. Main difference between samples *F-C-100(3)* and *F-C-100(5)* is the depth of microgrooves, much smaller in the former case (see Table S1).

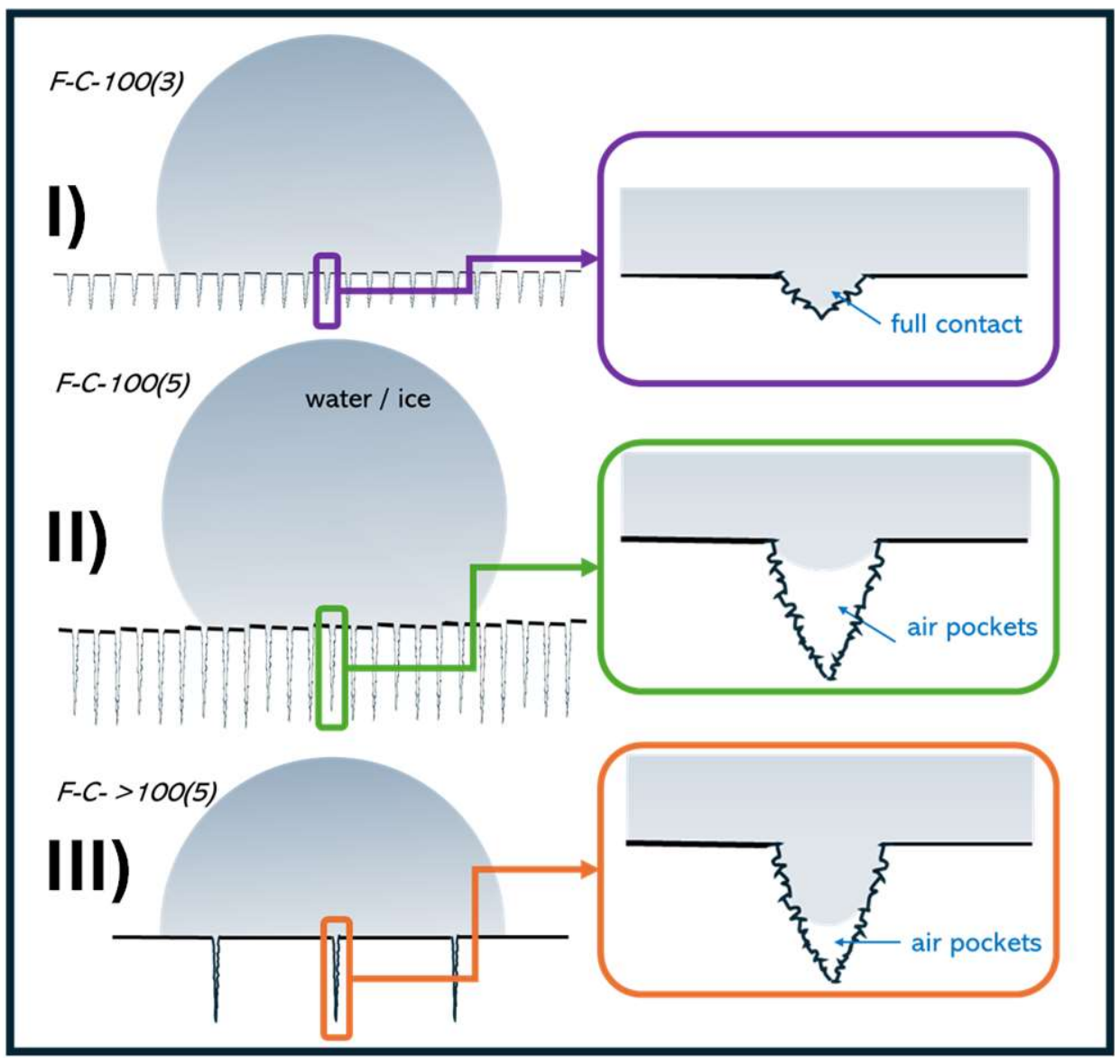

**Figure 5. Scheme representing the retention of air pockets inside the nanostructured microgrooves of *L-* and *C-* samples.** I-II) Wetting of sample *F-C-100(3)* and *F-C-100(5)* characterized by a high surface density of microgrooves with relatively small and high depths, respectively. In sample *F-C-100(3)* water fills completely the microgroove, but not in sample *F-C-100(5)* where air pockets remain at the bottom. III) Wetting of samples *F-C->100(5)*: water leaves air pockets inside the microgrooves but the apparent WCA is not so high because the surface fraction affected by the microgrooves is relatively smaller.

This makes that water can penetrate and fill completely the groove; the sample behaves as hydrophobic under the premises of a Wenzel wetting state. In sample *F-C-100(5),* the grooves are deeper, so the water does not fill in them completely and air pockets remain inside the grooves. This sample was highly hydrophobic as expected for a Cassie-Baxter wetting regime. Air pockets should also exist in the grooves of samples *F-C->100(5)*, but they do not reach a so high apparent WCA due to the large separation existing between microgrooves and the relatively small fraction of surface covered by microgrooves.[45] For example, the groove-free area fraction had values of 0.47 and 0.88 for samples *F-C-100(5)* and *F-C-500(5)*, respectively, which applying the typical C-B formula (1) renders $\theta^{CB}$ values of 139° and 123°, respectively, in good agreement with the experimental results in Figure 4 b).

### 3.4 Anti-fouling behaviour of transparent patterned glass surfaces

To assess the antifouling capacity of the patterned borosilicate samples, we have employed three different fouling fluid simulants: bovine serum as well as humic acid, and sodium alginate solutions. This combination, together with the response to different liquid polarities like water and diiodomethane, covers an ample range of variants such as the soil, physiological culture medium and seaweed. **Figure 6** gathers the CA to these fluids comparing the pristine glass reference and the laser patterned samples *C-100(5)* and *F-C-100(5)*.

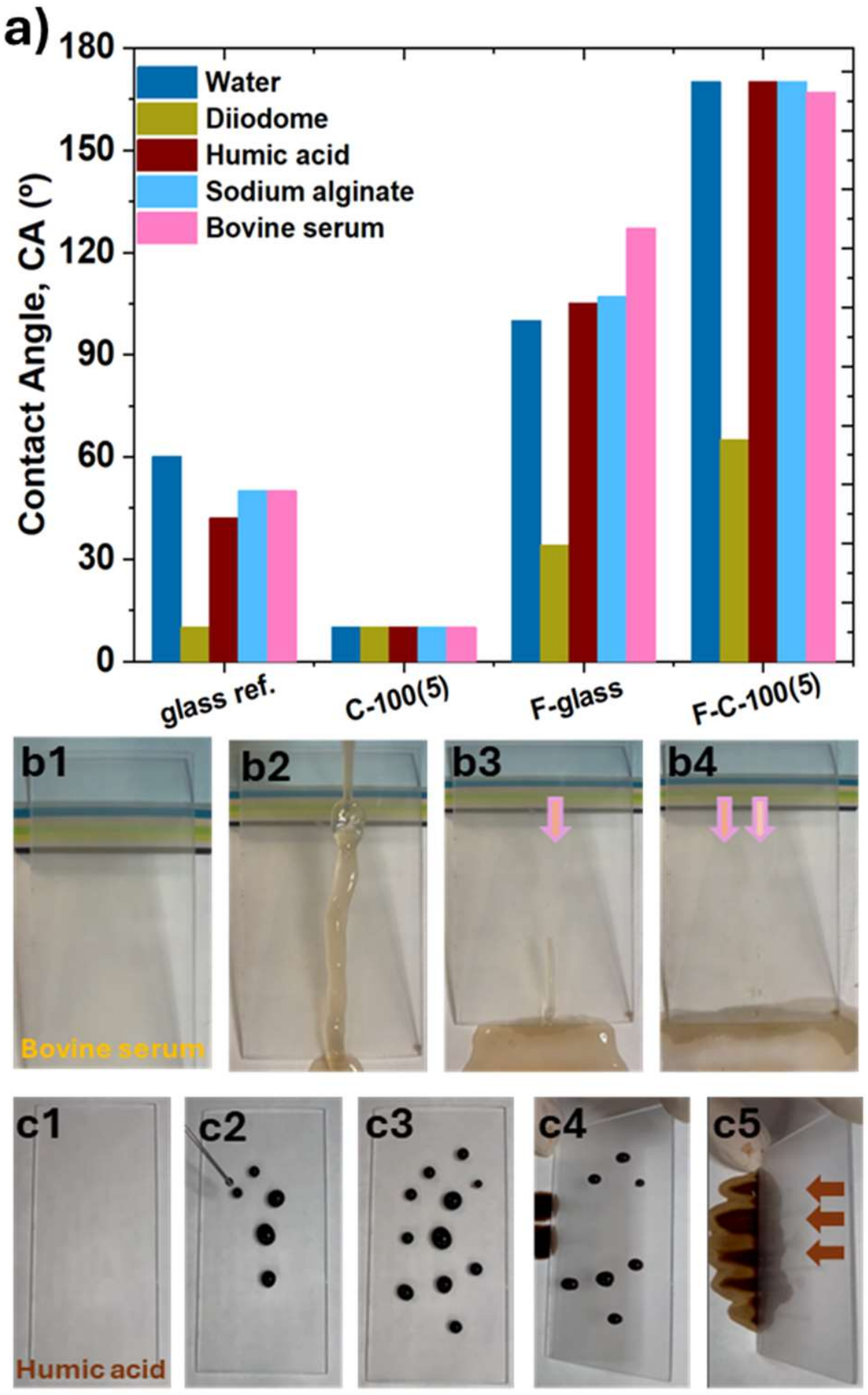

**Figure 6. Anti-fouling response of the studied transparent surfaces.** a) Wetting contact angles with polar, non-polar and organic simulant liquids for samples *glass ref., C-100(5), F-glass* and *F-C-100(5)*. Sequence of images of anti-fouling behavior using b) bovine serum (blood simulant), and c) humic acid (mud simulant) demonstrated by the *F-C-100(5)* sample. In image series b), the surface is exposed to a flow of bovine serum in a vertical position, with arrows indicating the process by which the liquid slides off the surface. In image c), humic acid droplets are placed on the sample surface in horizontal position and then tilted, with arrows indicating the direction of the droplets sliding.

On the one hand, non-fluorinated surfaces, as the reference and *C-100(5),* presented partial omniphilic behavior. On the other hand,

the CA with fouling fluid droplets drastically increased after molecular grafting. Figure 6 a) shows that sample *glass-ref* presents CAs around 115-125°. Superomniphobicity, with contact angles higher than 150°, was found for sample *F-C-100(5)*. The capacity of tailoring the sliding behaviour of liquids on patterned glass is further confirmed considering the RoAs of water, diiodomethane and fouling liquids on samples *C* and *F-C* (see Supporting Information Table S5), where values lower than 2° for samples *F-C-50(5)* and *F-C-100(5)* support total repellence and superomniphobicity. The repulsion of fouling liquids by this last sample was further confirmed by dynamic experiments, as those presented in Figures 6 b) and c). When a continuous stream of bovine serum was circulated over the surface in vertical position, (Figure 6 b2)) it was observed to pass by without dispersion until it left the surface without a trace (Figure 6 b3 and b4) in a demonstration of self-cleaning ability. This notable anti-fouling and self-cleaning capacity, also supported by the test where humic acid droplets were deposited on the *F-C-100(5)* surface (Figure 6 c2)) and after a slight inclination they quickly abandon it (Figure 6 c4) and c5)), is the result of a compromise between the friction and the adhesion forces of the sliding droplets on the surface.[64] In the case of lineal patterns, sliding on samples *F-L* was anisotropic depending on the direction of droplet travel with respect to the laser lines. Thus, 15 µl droplets of all simulants, except diiodomethane, slid when tilted at angles lower than 60º, and the droplet had to move in the direction of the laser micromachined lines, but remained retained up to 90º when tilted in the perpendicular direction (Table S4 in Supporting Information). This sliding behavior along the groove direction with more viscous and/or dense liquid droplets compared to water was particularly noticeable for samples with microgroove separations of 300 µm This behaviour is restrained for the particular pattern sizes of these samples since for groove distances smaller and greater than 50 - 100 µm, complete adherence (i.e. RoAs of 90°) was found, highlighting a petal-like wetting behaviour.[65]

### 3.5 Water condensation on patterned glass surfaces

The previous analysis has revealed that although most patterned samples present a high hydrophobic or superhydrophobic behaviour, the cohesive force of water droplets onto their surface is relatively high and preclude an effective sliding or rolling onto the surface. An exception to this trend was found for sample *F-C-100(5)* (c.f. Figure 4). This sample and sample *F-L-100(5)* have been chosen for water condensation experiments. Two scenarios were selected: an environmental chamber and an environmental electron scanning electron microscope. **Figure 7** displays a series of images taken at 2 °C and 98 % relative humidity (RH) in the environmental chamber with the samples in a vertical position. According to images in panels a1)-a4), taken after 2 to 60 min of exposure to the humid/cold conditions, the amount of water condensation on the surfaces followed the order: *glass ref*>> *F-L-100(5)* (lines in normal configuration respect to sample position) > *F-L-100(5)* (parallel configuration) > *F-C-100(5)*. In fact, a continuous layer of water was rapidly formed on the *glass-ref*. A rough estimation of the condensed volume was 1 µl/cm$^2$ after 11 min. Comparatively, much less water accumulated on the laser-patterned samples after equivalent time intervals. Furthermore, the condensed water in the line patterned surfaces adopted the form of isolated, irregularly shaped droplets as they grow, coalesce and touch the edges of the microgrooves .[46] Water condensation under 98% RH was slightly more pronounced on sample *F-L-100(5)* where large and irregular water droplets could be observed when the microgrooves were perpendicular oriented to the ground (c.f. Figure 7 b)). Conversely, water droplets were smaller and rather abundant for the microgrooves oriented parallel to the ground (compare Figure 7 b3) and c3)). However, in neither case there was an effective sliding of coalesced water droplets. This result, together with the RoAs measured at ambient conditions (Supporting Information Table S4) indicate that microgrooves in sample *F-L-100(5)* are not effective to drain the water accumulated on the surface under supersaturated humid conditions. This contrasts with reported results[38, 39] showing the usage of linear microgrooves in systems that usually combine hydrophilic and hydrophobic paths as effective drainage system to collect condensed water. Obvious differences between the patterns in these works and in our samples reside in the geometry of microgrooves, much deeper and wider in the former. This enables the growth and sliding of bigger droplets up to a few mm diameters. [38,39] . The limited growth of condensed droplets has been illustrated with a water condensation experiment in horizontal configuration comparing both *F-L-100(5)* and glass reference surfaces, presented in Supporting Information **Figure S7 a).**

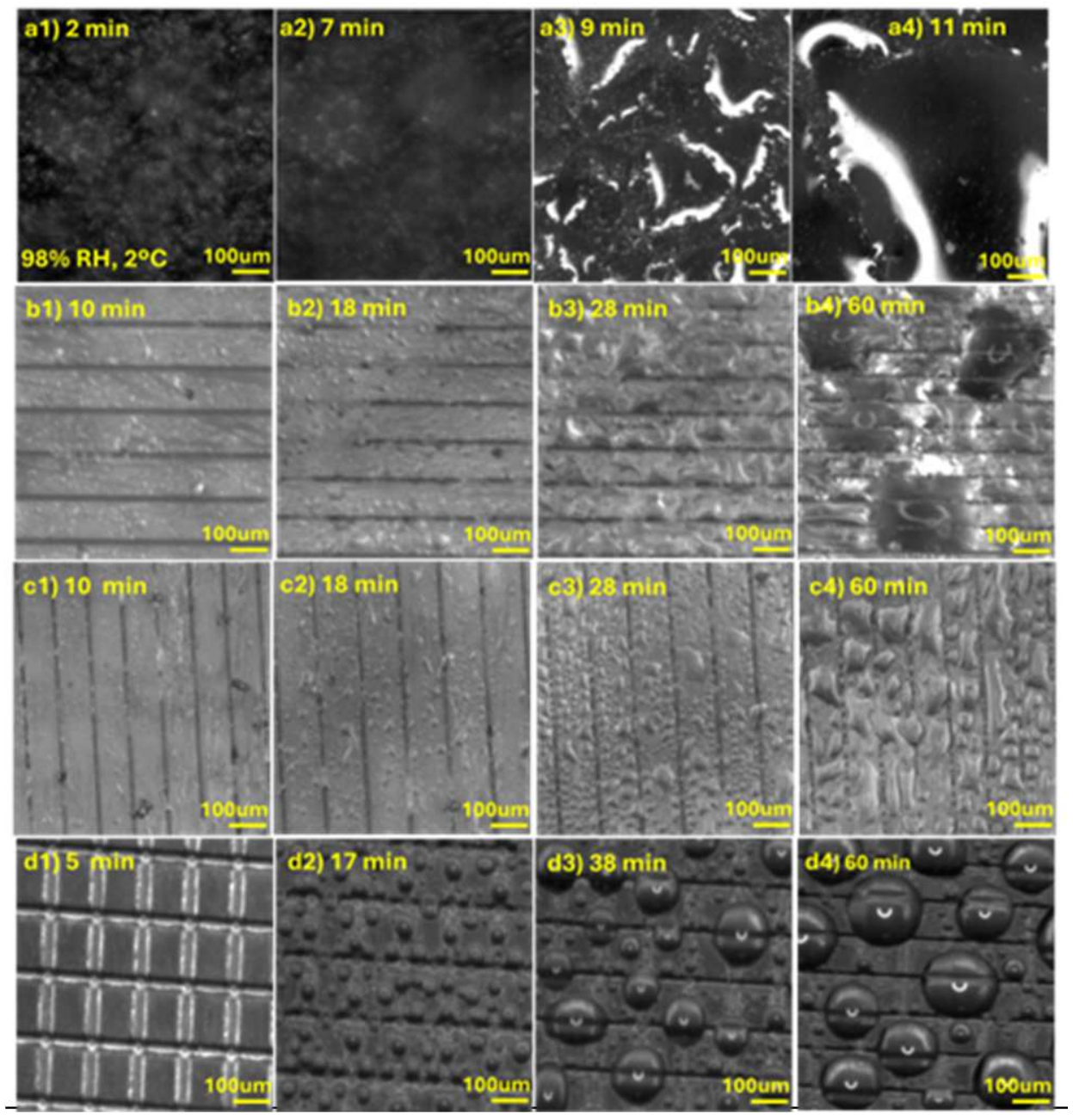

**Figure 7. Water condensation experiment conducted at the environmental chamber at 2 °C and 98 % RH for the samples placed in vertical position.** Snapshots were taken for the indicated times from 2 to 60 min for: a) *glass ref.*, b) and c) *F-L-100(5)* with the microgrooves placed in normal or parallel direction, respectively, and d) *F-C-100(5)*.

However, according to Figure 7 d), small spherical droplets only appeared on the *F-C(100)(5)* after a delay of 17 min of exposure. An estimation of water surface coverage from Figure 7 d3) (i.e., after 38 min) was 68 %, which is equivalent to 0.6 µl/cm$^2$. These results support that sample *F-C(100)(5)* effectively hurdles the condensation of water onto its surface. Interestingly, this condensation occurred through the formation of separated and small water droplets of similar size. In sample *F-C-100(5)*, the water condensation phenomenology was slightly different than in sample *F-L-100(5)*, albeit a small size of droplets in both cases. The initially detected condensed droplets had a size of approximately 100 µm in diameter ($5 \times 10^{-4}$ µl volume assuming a spherical shape); they coalesced after 60 min into large and regular droplets of around 300 µm diameter (estimated volume of $65 \times 10^{-3}$ µl) (Figure 7 d2) to Figure 7 d4)). This small size is rather stable over time and clearly precludes an effective sliding of droplets (note that larger 15 µl droplets dropped on the surface of this sample did slide off at angles as small as 2°, Supporting Information Table S5). It is known that at supersaturated humid conditions microdroplets which are condensed on superhydrophobic surfaces may transition from a Cassie-Baxter to a partially Wenzel state, this increasing surface adhesion and likely impeding sliding even if a critical size is not reached.[38] For patterned surfaces, depending on the depth of the grooves, newly condensed water vapor in the cold bottom regions may come into contact with existing droplets, promoting a transition in the wetting regime. [39] Although the small size and depth of the microgrooves necessary to maintain the transparency of our samples limit the formation of large droplets, these characteristics are beneficial for reducing the size of condensed water droplets and their overall quantity. They are also beneficial for controlling the coalescence of pinned microdroplets, creating clean, transparent areas around them, which is essential for anti-fogging functionalities.

In an attempt to further improve the understanding of the water condensation mechanisms on patterned samples, we examined the condensation process using an environmental scanning microscope, keeping the samples at 2 ºC while incrementally increasing the water vapor pressures. Selected results in **Figure 8** showcase the progress of condensation on samples *glass ref.*, *F-glass* and samples *C-100(5)* and *F-C-100(5)*. Vapor pressure was varied from 700 Pa up to 1200 Pa. On sample *glass ref.* (Figure 8 a)), liquid water condensates readily appeared on the surface, enlarged their size, and eventually formed continuous films of water at 1200 Pa. This behaviour is characteristic of a hydrophilic surface and reproduces well the macroscale condensation behaviour in air on a flat surface (cf. Figure 7 a)). On the hydrophobic flat sample *F-glass* (Figure 8 b)), condensation proceeded through the formation of small and quasi-spherical droplets that increase in size from 700 to 1000 Pa of water vapour pressure. Eventually, at 1200 Pa, large droplets coalesced with the neighbouring ones, although their diameter remained smaller than 100 µm.

Water condensation on the highly hydrophilic patterned sample *C-100(5)* revealed two well-differentiated regions (Figure 8 c)): initially, water accumulates in the form of a continuous film inside the microgrooves, followed by the formation of small droplets on top of the flat surface of the rectangular micropillars, in agreement with the expected effect of the roughness induced on a hydrophilic glass surface. Meanwhile, on the superhydrophobic *F-C-100(5)* (c.f., Figure 8 d)), formation of very small spherical droplets was only observed for water vapor pressures around 1000 Pa, i.e., above the water saturation pressure at the temperature of the sample. These water droplets formed preferentially on the edges of the micropillars and then on their flat surfaces. Since the effect of the distribution of water repellent functional groups has been found to be uniform over the entire surface of the glass, the delay in water condensation in the microgrooves must respond to the influence of the nanostructures generated inside them. Further increasing the vapor pressure to 1200 Pa produced an increase in the size of spherical microdroplets, although they were much smaller than those found on sample *F-glass* for the same conditions.

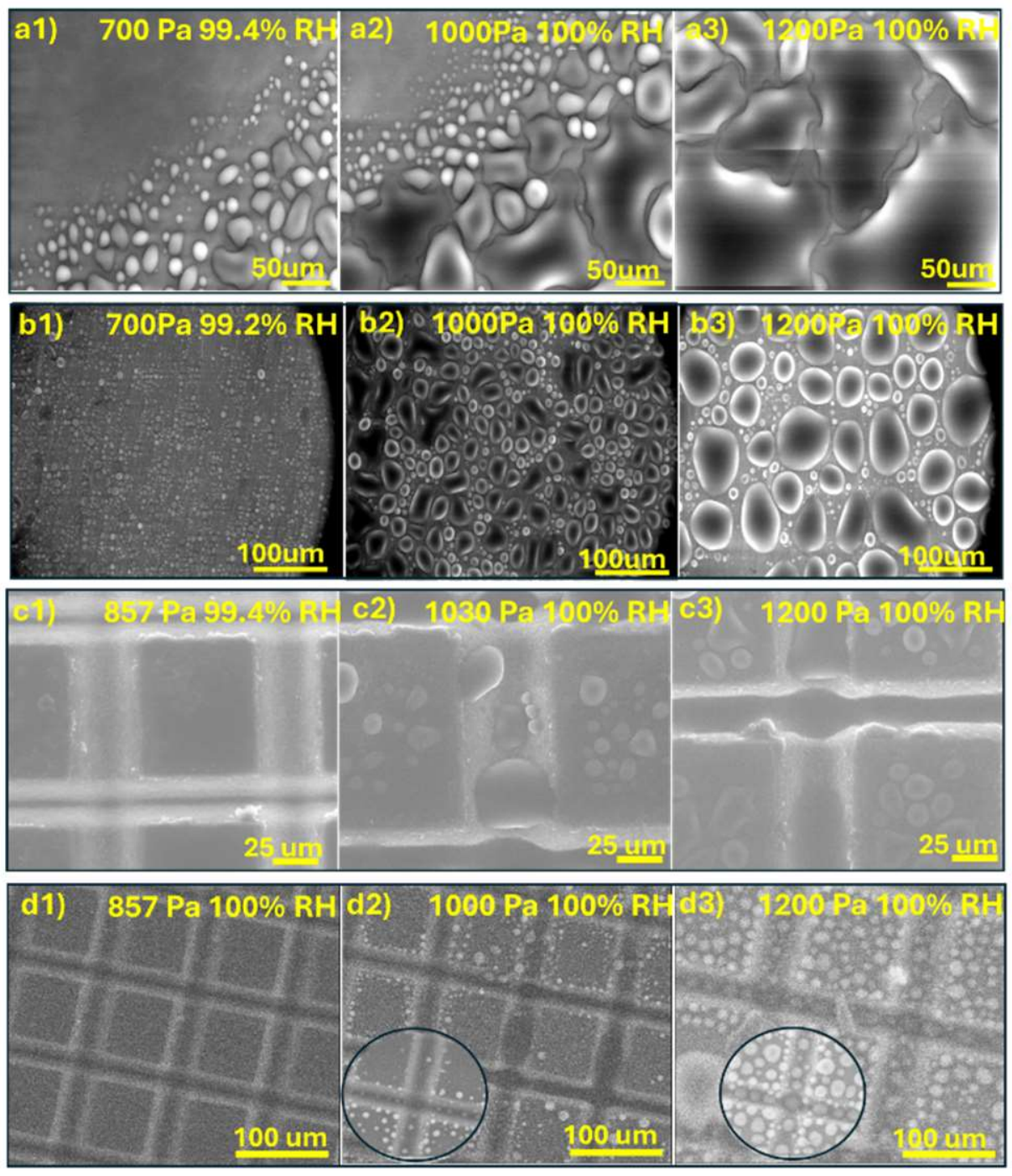

**Figure 8. Microscale water condensation experiment on pristine and laser-treated surfaces.** ESEM images of the evolution of water condensation at 2º C and increasing water vapor pressures onto samples a) *glass ref.*, b) *F-glass*, c) *C-100(5)*, and d) *F-C-100(5)*.

ESEM observations confirmed the experiments at atmospheric pressure ((Figure 7 d3)) and d4)) in the sense that functionalized microgrooves in sample *F-C-100(5)* prevent the coalescence of microdroplets into larger droplets. The fact that the size of microdroplets in the ESEM experiment was even smaller than in air must be attributed to the different experimental conditions in each case. Therefore, the ESEM experiments evidence that the low surface energy, its superhydrophobic character and the relatively small size of microgrooves in sample *F-C-100(5)* efficiently restrict the growth of water droplets on its surface, resulting in an outstanding anti-fogging glass functionality.

### 3.6 Anti-icing response of transparent patterned glass surfaces

In general, highly hydrophobic or superhydrophobic surfaces use to delay the formation of ice and reduce ice accretion.[41] However, superhydrophobicity relaying to hierarchical and multiscale roughness usually tends to yield high ice-adhesion strengths.[21] Indeed, a perfect solution for anti-icing surfaces is yet to be defined, and it is extremely challenging for transparent surfaces. In this context, we have evaluated the anti-icing properties of the surfaces, considering both the Freezing Delay Time (FDT), as shown in Figures 4 a) and b), and the ice adhesion strengths, evaluated by a pull-off and push-off experimental setups (**Figure 9**).The longest FDT values around 100-110 min, were found for samples *F-L-10(5) and F-L-25(5)*, where microgrooves overlap forming a continuous nanostructured surface (c.f., Figures 2 a) and b)). Then, FDT values progressively decreased parallel with the increase of the distance between microgrooves. These followed the evolution of WCAs, reaching a minimum value of 38 min for sample *F-L-500(5)*. However, this value was still higher than that found on the flat sample *F-glass-ref*. All *F-C* samples presented FDT values that follow the evolution of WCAs (Figure 4 b)). Outstanding FDT values of 180 and 168 min were found for samples *F-C-100(5)* and *F-C-50(5)* , respectively. Compared to current transparent anti-icing materials mostly based on the implementation of coatings,[66,67] the FDT values of *F-L* and *F-C* samples are one order of magnitude higher, demonstrating the effective anti-icing potential of the laser patterning/grafting procedure. Remarkably, for five-laser-scans samples, the FDT values follow the percentage of surface laser affected area.[68] This supports the fact that, either the existence of air pockets limiting the contact points between water/ice and glass (see Figure 5), or the micro-roughness of the (limited) contact zones inside microgrooves, restricts the formation of sufficiently large ice nuclei.[69] Moreover, according to our model in Figure 5, it can be assumed that the greater the number of microgrooves under the contact interface with the supercooled water droplet, the lower the conductive heat loss,[70] as the air pockets would act as effective thermal insulators as a consequence of their low thermal conductivity[71] (~25 $mWm^{-1}K^{-1}$ compared to ~1 $Wm^{-1}K^{-1}$ typical of borosilicate glass[72]). Worsening of the freezing delay as the distance between the microgrooves in *C-* patterned five scan samples might be inversely related to the number of grooves in the contact area with the supercooled water droplet. For a 2 μl supercooled water droplet (note the approximation: the actual number of grooves in contact with the droplet will vary with the contact area and this depends on the value of the apparent WCA), the number of grooves beneath the water decreases from 339 to 15 for *C-100* to *C-500* samples, respectively, implying a variation ranging from 53% to 12% in the number of microgrooves and, therefore, air pockets content below the water droplet interface.

Ice adhesion strength is another key parameter to characterize the anti-icing capacity of surfaces.[42] This property is significantly influenced by the wetting capacity of surfaces and their planarity. Rough surfaces typically lead to high adhesion strengths due to interlocking between the ice and the roughness features at the surface. [10,41] The analysis of the ice adhesion strength of the studied samples confirms these basic tendencies with some specificities that deserve attention. The values reported in Figures 9 a) and b) reveal that sample *F-C-100(5)* presented a minimum ice adhesion strength in pulling configuration outstandingly, almost similar to the corresponding to the Teflon sample used as reference (~23 kPa), closely followed by samples *F-C-50(5) and F-C-200(5)*. Moreover, this non-stick behaviour against ice has been confirmed also with a push-off ice adhesion test performed in an icing wind tunnel at -10 ºC (Figure 9 c)). A pushing ice adhesion value similar to that of Teflon reference has been also recorded for the *F-C-100(5)* surface, being 3.6 or 5.6 times lower than that corresponding to the untreated or only laser treated glass surfaces, respectively. According to the hypothesis represented in Figure 5, in samples *F-L-* and *F-C-* large air pockets kept in the microgrooves will remain beneath ice particles upon water freezing. This will make that part of the ice-glass is substituted by an ice-air interface with a significant reduction of the ice adhesion stress. As expected, the effect relaxes for the samples where the grooves are more separated. Interestingly, sample *F-L-100(5)* with an adhesion strength of 41 kPa also presented the minimum ice adhesion value among all *F-L* samples (in this case, not correlated with the more hydrophobic line patterned surface).

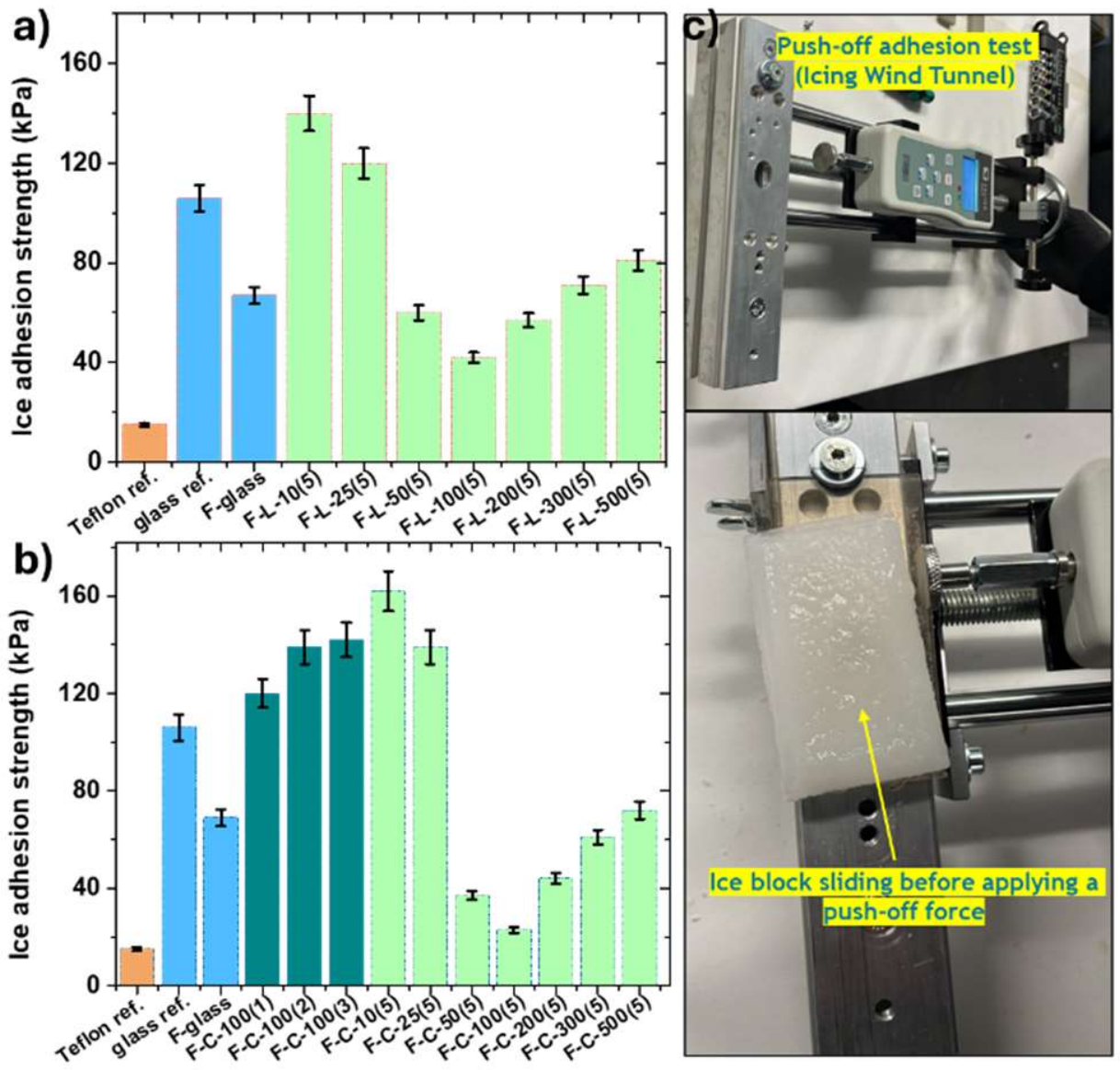

**Figure 9**. **Ice adhesion strength tests.** Ice adhesion strength evaluation by pull-off experiments at -13 ºC for a) *F-L-* and b) *F-C-* samples compared to flat reference glass and Teflon; c) Pictures of the push-off ice adhesion test showing the horizontal configuration of the dynamometer (top), and a mixed glaze ice block, formed at -10 ºC in the icing wind tunnel, sliding smoothly on top of the superhydrophobic *F-C-100(5)* surface after a minimal lateral pushing contact.

Furthermore, for both *F-L* and *F-C* samples, the adhesion strength dramatically increased for those samples with nominal groove separations of 10 and 25 μm, where microgrooves actually overlap, and the surface depicts a continuous and highly rough topography at the micro- and nano- scales to induce ice mechanical interlocking. Significantly, samples *F-C-100(1)-(3)* also presented a relatively high adhesion strength, despite their 100 μm microgroove separation. These results inversely mirror the apparent WCA trend determined for these samples (c.f. Figure 4), a feature that we relate with insufficient microgroove depth to ensure air pocket retention (see Figure 5-I and Figures S1 and S3 in Supporting Information).

In practical terms, the low ice adhesion strength found for samples *F-C-100(5)* and *F-L-100(5)* and their relatively high transparency in the visible region have interesting outdoor applications, as proven in the "open air" experiment to test the adhesion of snow onto these glasses. The experiment was performed at -4° C facing sample *F-L-100(5)* directly to falling snow from outside for more than 20 min (see video S1 and Figure S7 b) in Supporting Information). Neither snowflakes nor supercooled water droplets could be fixed on *F-L-100(5)* surface but passed by without coming into contact with it (though traces of condensed water could be seen on its back due to its transparency). Finally, another proof-of-concept experiment was carried out to check the anti-icing capacity of the highly transparent micropatterned glass samples with an infused fluorinated lubricant (i.e., samples *K-L-25(5)* and *K-C-100(5))*. It was found that sample *K-C-100(5)* exhibited the minimum ice adhesion strength (i.e., 17 kPa) from all investigated samples, supporting the potential relevance for outdoors applications demanding visibility (**Figure S8** in Supporting Information).

## 4 Conclusions

This work reports on the development of a procedure consisting of the laser patterning treatment followed by the surface molecular functionalization of glass samples to confer them superhydrophobicity, anti-icing and anti-fogging capacities, as well as anti-fouling properties, all without significantly altering their transparency. Glass is a hydrophilic material prone to become covered by water (e.g., by condensation), accrete ice and retain fouling liquids. On the other hand, glass is a fragile material that hardly withstands localized thermal stress without breaking. The surface processing technology developed in this work tries to cope with all these restrictions and requirements. The use of a femtosecond laser and the application of a precise multi-scan protocol to produce microgroove patterns on the surface of glass proved to successfully address these challenges. Different linear and rectangular microgroove patterns were manufactured, keeping defined widths for the carved lines but changing the distance between them. The best trade-off situation to simultaneously bestow the surfaces with an outstanding multifunctional response toward liquids and ice while keeping transparency is with microgroove patterns formed by five time-scanned lines separated by 100 microns. This configuration appears to be particularly well-suited to generate superhydrophobicity, to reduce ice adhesion, and to prevent condensation of water. These functions can be interpreted by considering the interplay effects involving the interaction of liquids with the flat surface regions on glass, the extremely rough surfaces in the microgrooves, and the influence of capillary forces in the microgroove channels. It has been evidenced that although the width/height ratio of the linear microgrooves and the wetting properties of the untreated surface zones do not induce an effective sliding of water droplets at room conditions, the tailored topographies are particularly well suited to: i) prevent the formation of large water droplets upon condensation, ii) reduce the ice adhesion force, iii) prevent the accumulation of snow on these surfaces. It has been also found that the multifunctional response of the patterned glass can be tailored within a wide parametric range to achieve an effective control over the apparent water contact angle, the antifouling behavior and the ice adhesion forces, while keeping the transparency in the visible region within manageable limits. A proof-of-concept experiment has also shown the possibility to infuse liquids in the microgroove structure of the glass to manufacture SLIPS with high transparency and notable anti-icing properties. It is expected that these breakthroughs will set up the basis for future glass technologies, where preservation of transparency is made compatible with the multifunctional capabilities required for many outdoor applications.

## ASSOCIATED CONTENT

**Supporting Information**.

**Figure S1**. Confocal topographical profiles for the *C-100* pattern with a) 1 b) 2 c)3 and d) 5 repetitions. **Figure S2**. SEM images for line and cross laser patterns with larger distances between them: a) *L-200*, b) *L-300*, c) *L-500*, d) *C-200*, and e) *C-500* after 5 repetitions. **Figure S3**. SEM images at different magnifications of cross laser microgrooves with fixed distance between lines (*C-100* sample), and in function of repetitions: a) one laser scan, b) two laser scans, and c) 3 overlapping laser times. Last images correspond to the interface between a microgroove and the flat surface. **Figure S4**. XPS analysis of the glass surface before and after molecular grafting in both glass reference and *C-100* laser pattern: a) survey, b) C 1s, c) Si 2p and d) O 1s spectra, and of two point laser treated glass surface regions separated 50 µm corresponding to the glass plateau and the bottom of the microgroove: e) F 1s, f) C 1s, g) Si 2p and h) O 1s. **Figure S5**. Optical properties before and after molecular grafting of *L- 100(5)* and *C-100(5)* patterned surfaces. **Figure S6**. Optical transmittance of a) sample *F-L-25(5)* and b) sample *F-C-100(5)* before and after infiltration with the Krytox lubricant liquid (samples *K-L-25(5)* and *K-C-100(5)*). **Figure S7.** a) Sequence of images of a water condensation test in a controlled environmental chamber at 2°C and 48% RH and increasing times for a reference sample (top) and *F-L-100(5)* sample (bottom) including two respective zoomed regions of the last time event in horizontal configuration; b) sequence of images of a real "open air" experiment with the *F-L-100(5)* sample at -4 °C for different snow exposure times. **Figure S8**. a) Water contact angles, and freezing delay times, and b) ice adhesion measurements, for SLIP surfaces after the infusion of Krytox lubricant. Comparison of samples *F-L-25(5), F-C-100(5)* with samples *K-L-25(5), K-C-100(5)*. **Table S1**. Microgroove widths and depths of patterns obtained by different laser treatment repetitions for the *C-100* sample. **Table S2**. Roughness parameters deduced from the confocal microscopy measurements over a 283 µm x 334 µm scanned area (including at least a microgroove at larger separations) for different overlapping laser cycles, different patterns and different distances between microgrooves. **Table S3**. Atomic concentrations at the surface of laser treated and untreated glass samples compared to the reference before and after molecular grafting. **Table S4.** Rolling-off angle values of 15 µl droplets of different liquids on *L* and *F-L* samples as function of line separation before and after molecular grafting measured in parallel and normal (in brackets) directions with respect to the laser microgrooves. **Table S5**. Rolling-off angle values of 15 µl droplets of different liquids on samples C and *F-C* as function of the microgroove separation. **Video S1**. Real outdoor experiment with the *F-L-100(5)* sample at -4 °C.

"This material is available free of charge via the Internet at http://pubs.acs.org."

## ACKNOWLEDGMENT

The authors acknowledge the projects PID2022-143120OB-I00, TED2021-130916B-I00 funded by MCIN/AEI/10.13039/501100011033 and by "ERDF (FEDER)" A way of making Europe, Fondos NextgenerationEU and Plan de Recuperación, Transformación y Resiliencia", and Project ANGSTROM ( Joint Transnational Call 2023 of M-ERA.NET 3, (Horizon 2020 grant agreement No 958174). Also, the authors acknowledge financial support from the CNRS-CEA "METSA" French network (FR CNRS 3507) for the access to ESEM microscope, as well as the IBIMA Plataforma BIONAND for the ESEM service and the Determination of Surface Wetting, scientific-technical service of the Institute of Materials Science of Seville, funded by Junta de Andalucia (IE19_142 USE I+D+I Infrastructure and Equipment 2019-PAIDI2020 European Regional Development Fund). Niklas Kandelin and Raul Kanter of Tampere University are acknowledged for their assistance in icing research. This research work is funded by the EU H2020 program under grant agreement 899352 (FETOPEN-01-2018-2019-2020 – SOUNDofICE).

## Supporting Information

# Laser Patterning of Superhydrophobic transparent glass surfaces with anti-fogging and anti-icing applications

Laura Montes-Montañez,[1] Fernando Nuñez-Galvez,[1] Melania Sanchez-Villa,[1] Luis A. Angurel,[2] Raul Kanter,[3] Niklas Kandelin,[3] Heli Koivuluoto,[3] German F. de la Fuente,[2] Víctor Trillaud,[3] Philippe Steyer,[3] Karine Masenelli-Varlot,[3] Francisco J. Palomares,[4] Ana Borrás,[1] Agustín R. González-Elipe,[1] Carmen López-Santos,[1,5] * Víctor Rico,[1]*

1 Nanotechnology on Surfaces and Plasma Lab, Institute of Materials Science of Seville (US-CSIC) Americo Vespucio 49, 41092 Seville (Spain)

2 Institute of Nanomaterials and Nanoscience of Aragon (UNIZAR-CSIC) María de Luna 3, 50018 Zaragoza (Spain)

3 INSA Lyon, Universite Claude Bernard Lyon 1, CNRS, MATEIS, UMR5510, Villeurbanne, 69621, France

3 Faculty of Engineering and Natural Sciences, Materials Sciences and Environmental Engineering, Tampere University, FI-33014 Tampere (Finland)

4 Institute of Materials Science of Madrid (CSIC), Sor Juana Inés de la Cruz 3, 28049-Madrid (Spain)

5 Departamento de Física Aplicada I, Escuela Politécnica Superior, Universidad de Sevilla, Virgen de África, 41011 Sevilla (Spain)
*mclopez13@us.es (C.L.S.); *victor@icmse.csic.es (V.R.)

**Table S1**. Microgroove widths and depths of patterns obtained by different laser treatment repetitions for the C-100 sample.

| Patterned surfaces | Scan parallel to short axis laser, b | | Scan parallel to long axis laser, a | | Width at half height in-depth (μm) |
|---|---|---|---|---|---|
| | Width (μm) | Depth (μm) | Width (μm) | Depth (μm) | |
| **C-100(1)** | 24 | 2 | 20 | 3 | 18.4 |
| **C-100(2)** | 25 | 5 | 23 | 5 | 17.1 |
| **C-100(3)** | 29 | 8 | 25 | 8 | 15.3 |
| **C-100(5)** | 37 | 10 | 26 | 15 | 13.8 |

**Supporting information S1.-** ***Topographical profiles of C-100 microgrooves as a function of the number of laser scans.***

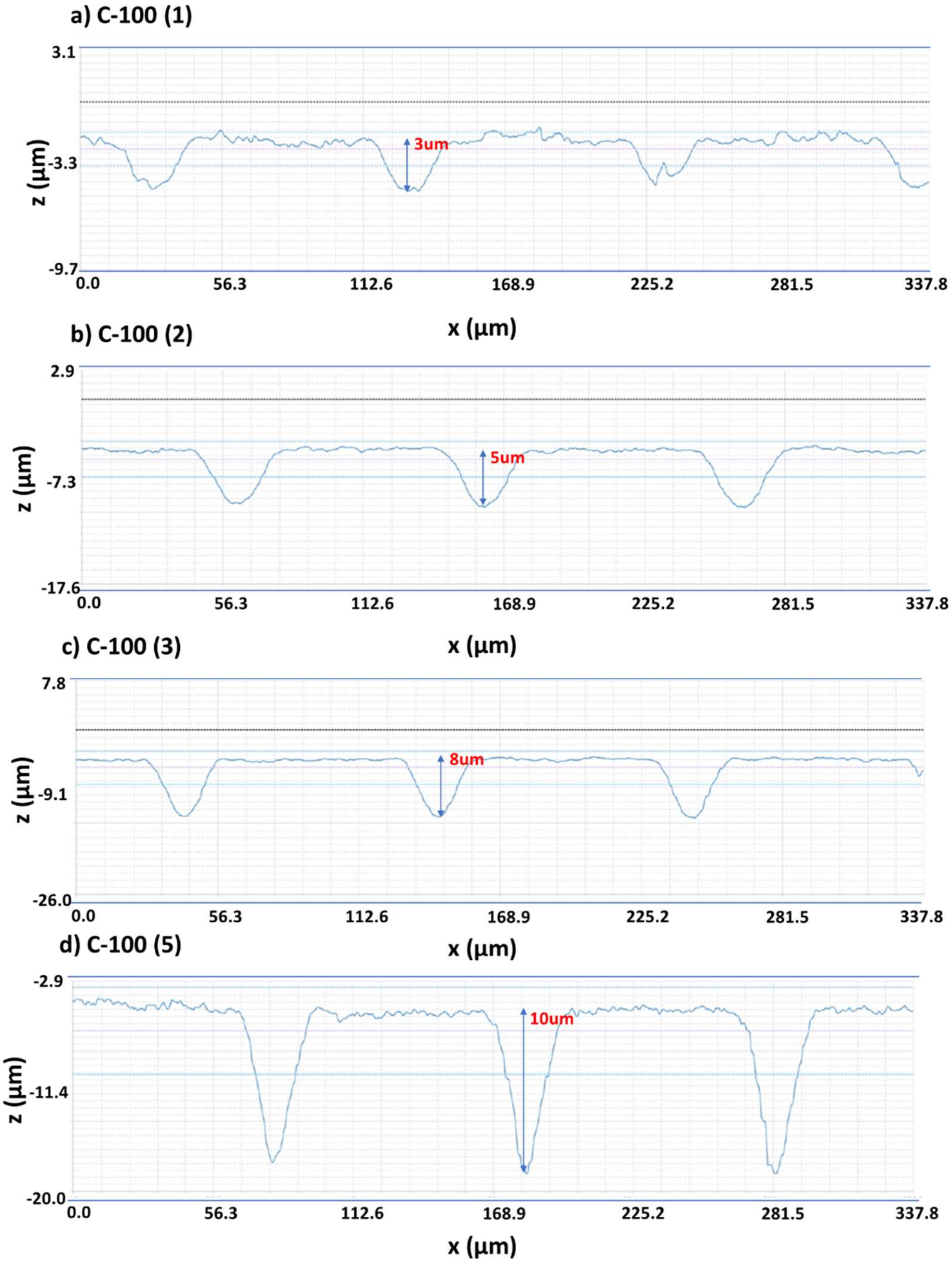

**Figure S1**. Confocal topographical profiles for the *C-100* pattern with a) 1 b) 2 c)3 and d) 5 repetitions.

**Supporting information S2.-** ***Surface morphology determined by SEM of the patterned samples after 5 laser scans.***

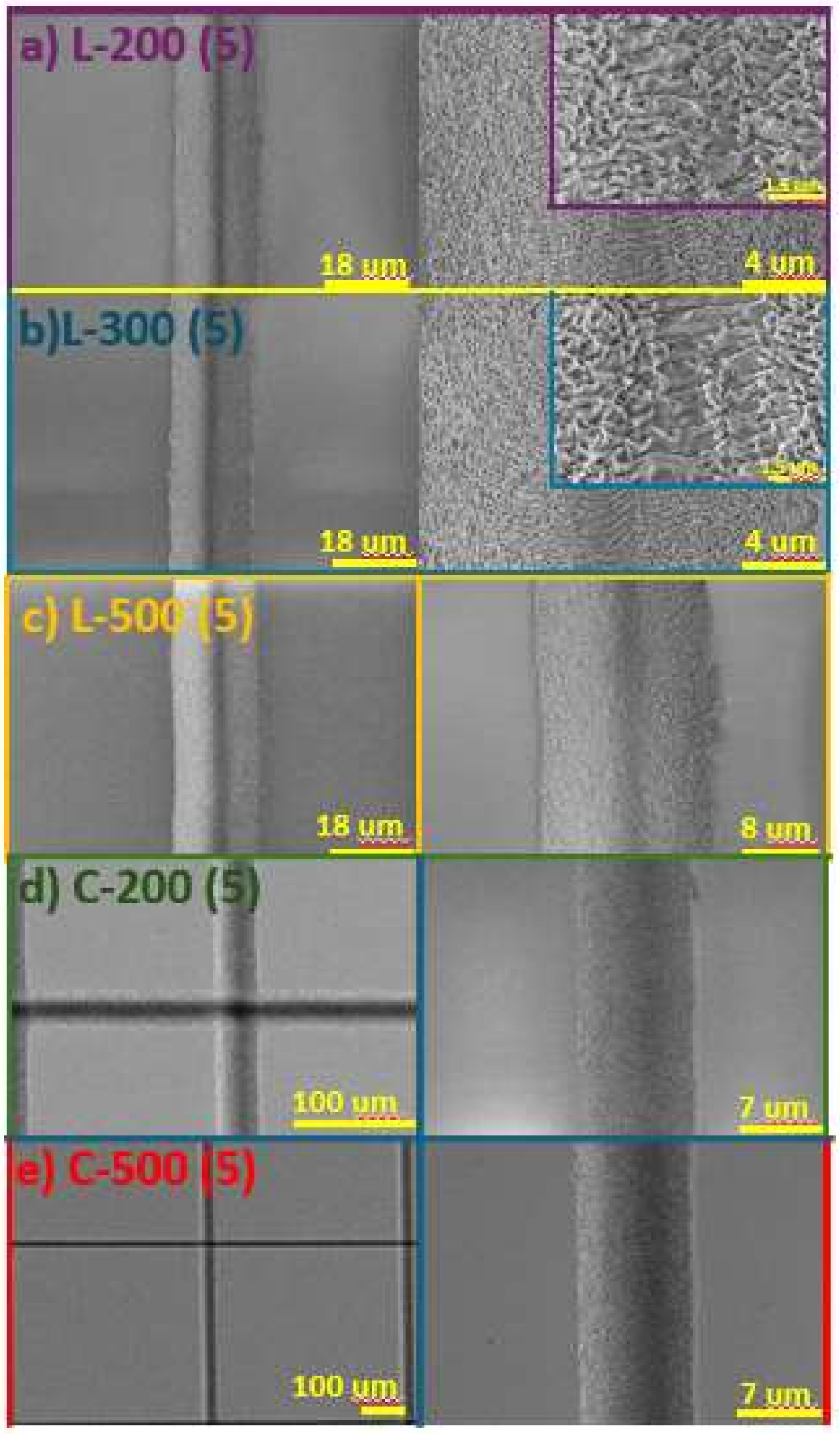

**Figure S2**. SEM images for line and cross laser patterns with larger distances between them: a) *L-200,* b) *L-300,* c) *L-500,* d) *C-200,* and e) *C-500* after 5 repetitions.

The images at higher magnification taken on the microgrooves show that both their walls and their valley regions depict a highly rough topography where very small and homogeneous nanostructured features can be devised.

**Supporting information S3.-** ***Surface morphology determined by SEM of the patterned samples depending on the number of laser scans.***

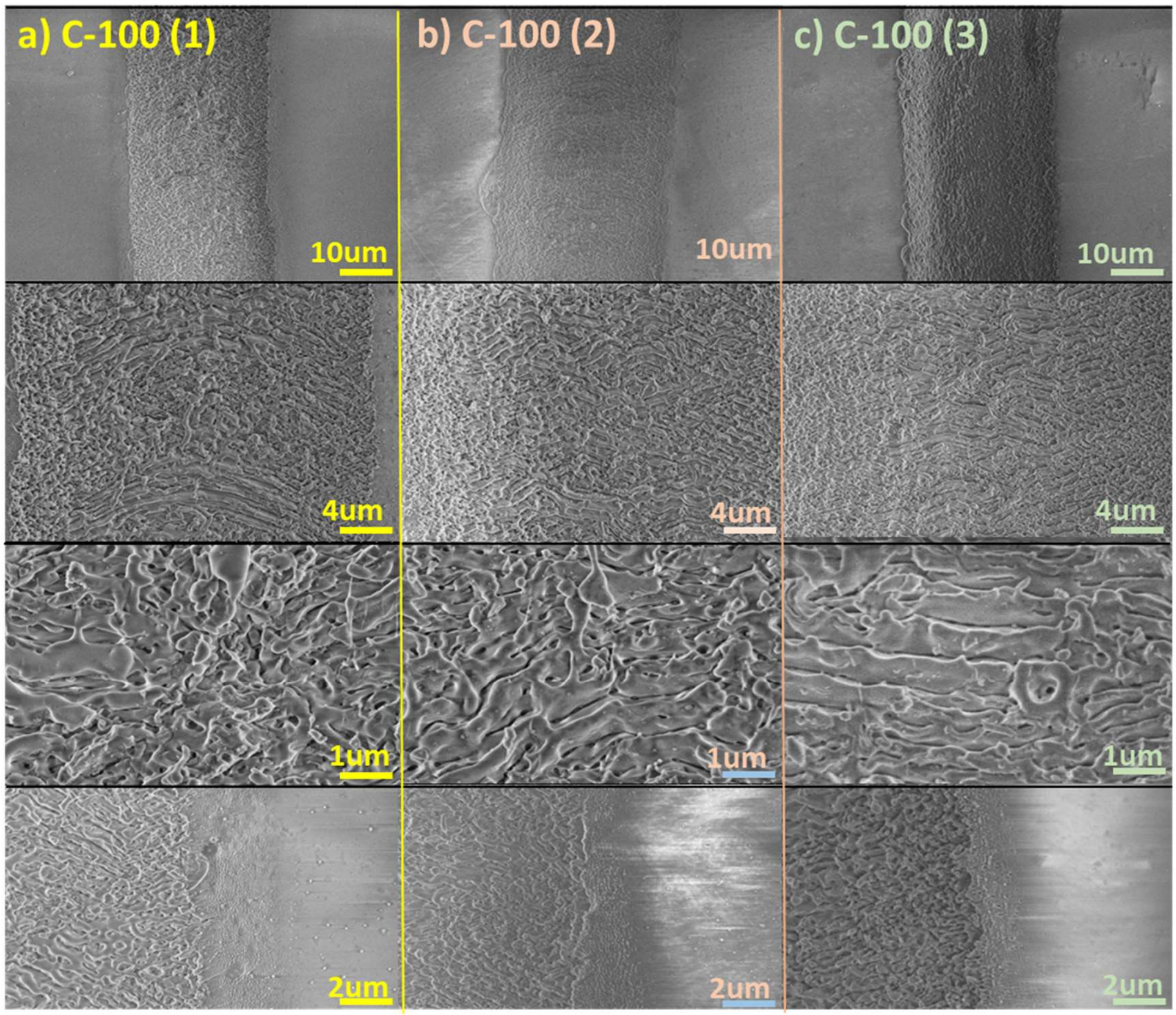

**Figure S3**. SEM images at different magnifications of cross laser microgrooves with fixed distance between lines (*C-100* sample), and in function of repetitions: a) one laser scan, b) two laser scans, and c) 3 overlapping laser times. Last images correspond to the interface between a microgroove and the flat surface.

The images at higher magnification show that the surface topography of the microgrooves only varies slightly with the number of scans. In all cases the surface in the microgroove zone (walls and valley) is rough and appears formed by nanostructured features of a rather similar type that suggest a certain melting of the glass during the ablation process.

## Supporting information S4.- Surface roughness of laser treated sample

**Table S2**. Roughness parameters deduced from the confocal microscopy measurements over a 283 μm x 334 μm scanned area (including at least a microgroove at larger separations) for different overlapping laser cycles, different patterns and different distances between microgrooves.

| | $S_a$ (μm) | $S_q$ (μm) |
|---|---|---|
| glass. reference | 0.50 | 0.54 |
| Line-distances | | |
| *L-10(5)* | 0.50 | 0.63 |
| *L-25(5)* | 3.40 | 3.89 |
| *L-50(5)* | 3.60 | 4.00 |
| *L-100(5)* | 4.00 | 4.29 |
| *L-200(5)* | 1.09 | 2.43 |
| *L-300(5)* | 1.00 | 2.20 |
| *L-500(5)* | 1.00 | 2.30 |
| Cross- distances | | |
| *C-10(5)* | 0.53 | 0.63 |
| *C-25(5)* | 2.90 | 3.50 |
| *C-50(5)* | 2.00 | 2.25 |
| *C-100(5)* | 3.00 | 3.20 |
| *C-200(5)* | 2.26 | 3.40 |
| *C-300(5)* | 1.5 | 2.0 |
| *C-500(5)* | 2.30 | 3.20 |
| Overlapping laser cycles | | |
| *C-100(1)* | 1.14 | 1.35 |
| *C-100(2)* | 2.32 | 2.70 |
| *C-100(3)* | 2.50 | 3.00 |
| *C-100(5)* | 3.00 | 3.20 |

**Supporting information S5.-** ***XPS analysis of the surface of samples after laser patterning and molecular grafting.***

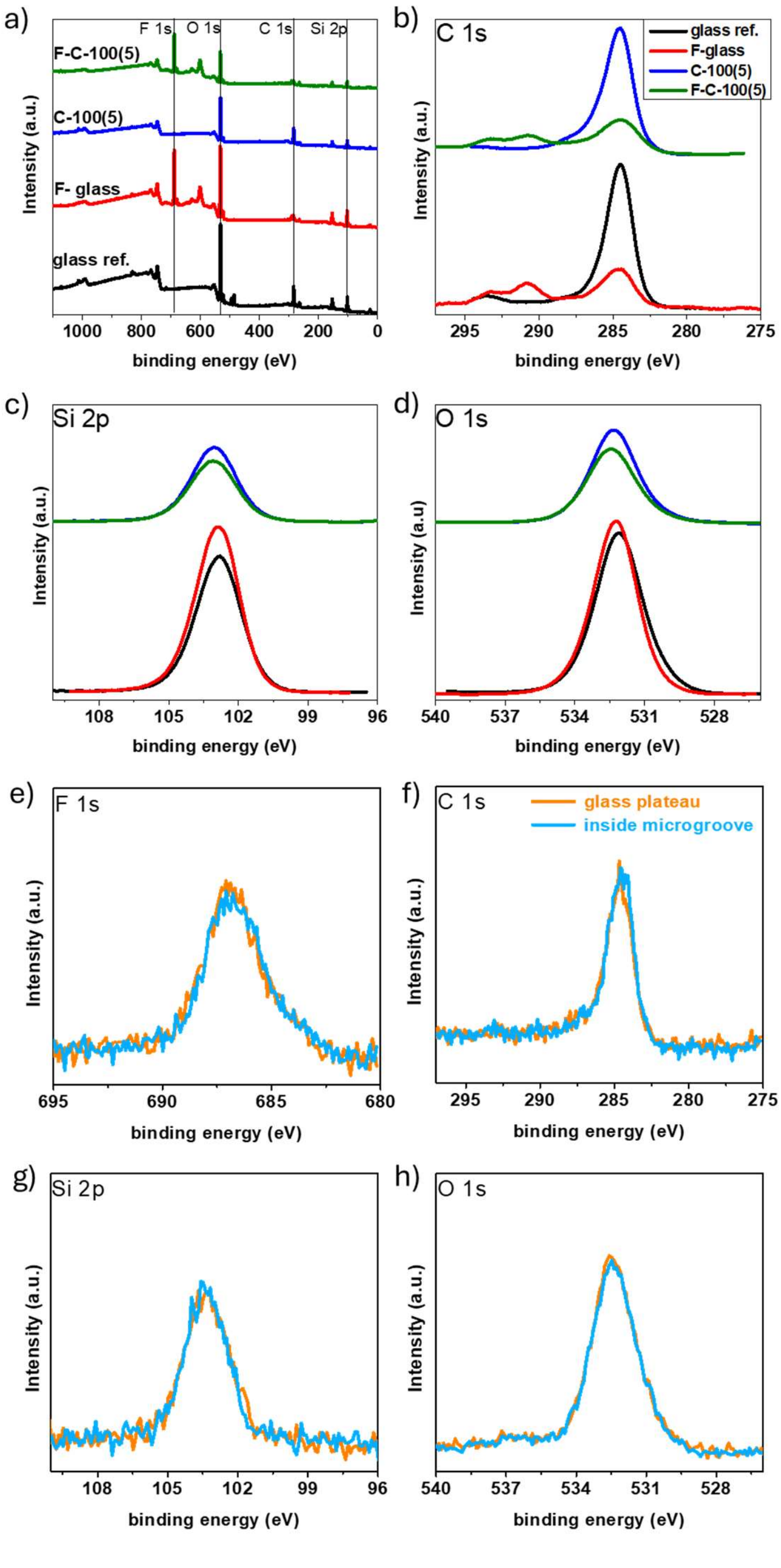

**Figure S4**. XPS analysis of the glass surface before and after molecular grafting in both glass reference and *C-100* laser pattern: a) survey, b) C 1s, c) Si 2p and d) O 1s spectra. Comparison of the spectra obtained on two point

laser treated glass surface regions separated 50 μm corresponding to the glass plateau and the bottom of the microgroove: e) F 1s, f) C 1s, g) Si 2p and h) O 1s.

The survey spectra clearly reveal the formation of a new F peak in the grafted samples. In the high resolution spectra of the C 1s level, two new peaks appear at binding energy higher than 290 eV attributed to carbon forming part of -$CF_2$- and -$CF_3$ groups of the PFOTES molecule. This assessment is confirmed by the atomic percentages determined for these samples, shown in the Table S3. Moreover, chemical composition of the laser treated glass after the fluorinated grafting uniform regardless of the XPS surface analysis location.

**Table S3**. Atomic concentrations at the surface of laser treated and untreated glass samples compared to the reference before and after molecular grafting.

| Atomic percentage (%) | C | O | Si | F |
|---|---|---|---|---|
| *glass ref.* | 28.8 | 46.3 | 24.9 | - |
| *F-glass* | 17.9 | 37.1 | 23.6 | 21.4 |
| *C-100(5)* | 41.3 | 39.1 | 19.6 | - |
| *F-C- 100(5)* | 25.1 | 30.2 | 16.9 | 27.8 |

**Supporting information S6.-** ***Optical transmission of patterned samples before and after molecular grafting***

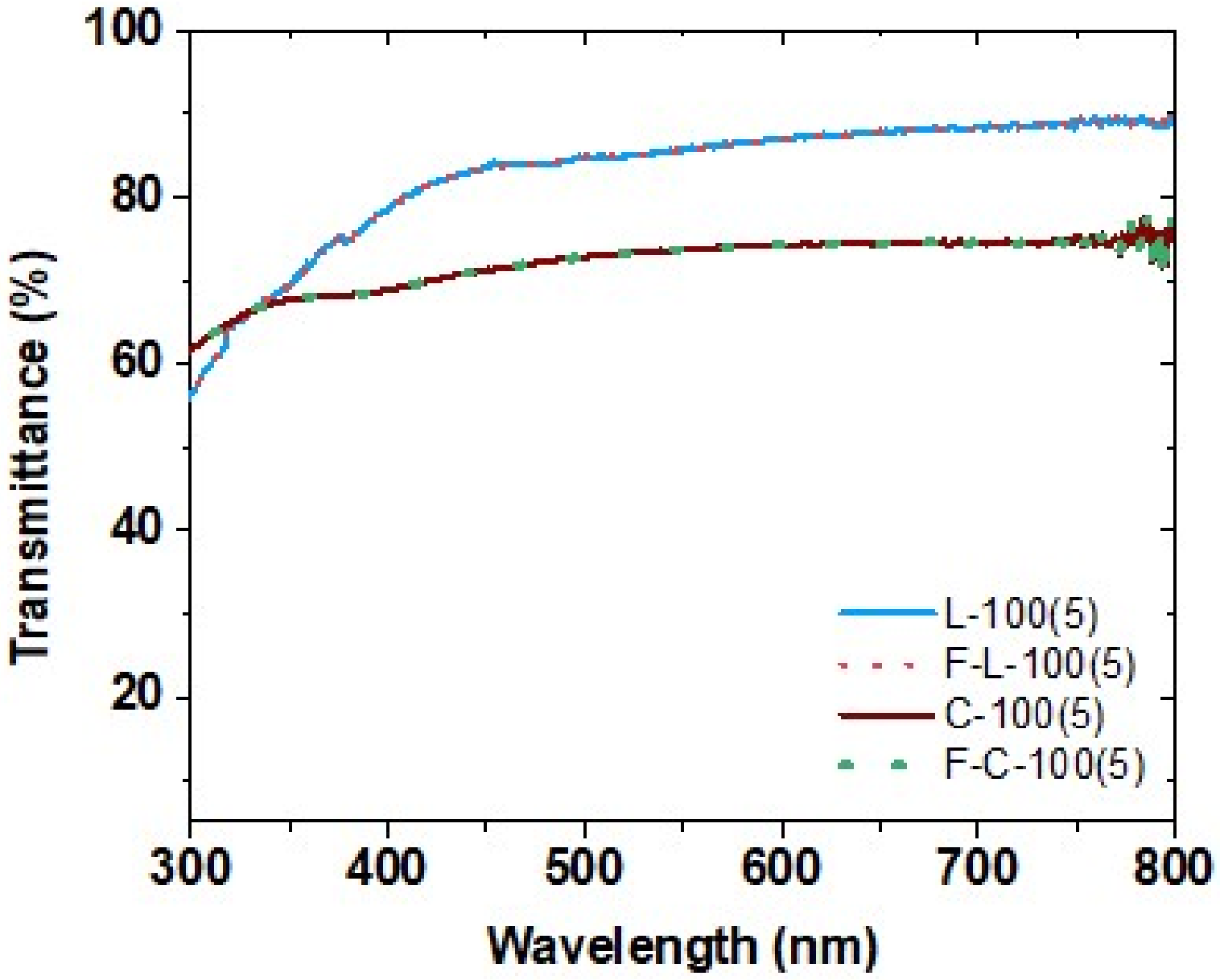

**Figure S5**. Optical properties before and after molecular grafting treatment of *L- 100(5)* and *C-100(5)* patterned surfaces.

No change in optical properties can be detected when comparing the optical transmission spectra of patterned samples before and after molecular grafting.

**Supporting information** S7.- ***Optical properties of SLIPS samples***

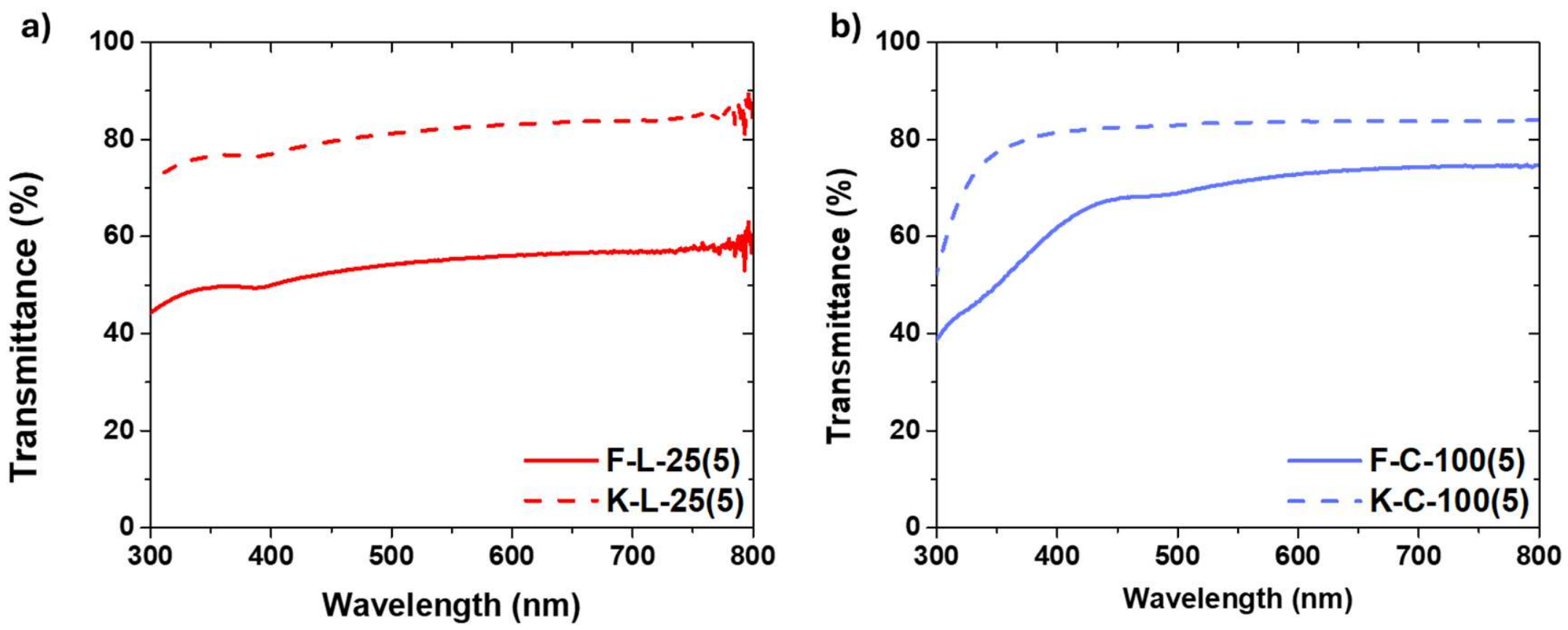

**Figure S6**. Optical transmittance of a) sample *F-L-25(5)* and b) sample *F-C-100(5)* before and after infiltration with the Krytox lubricant liquid (samples *K-L-25(5)* and *K-C-100(5)*).

The spectra show a net increase in optical transmission for all regions of the explored electromagnetic spectrum.

**Supporting information S8.-** ***Rolling-off angles (RoA) of different biological fluid simulant droplets on samples L and F-L.***

**Table S4.** Rolling-off angle values of 15 μl droplets of different liquids on *L* and *F-L* samples as function of line separation before and after molecular grafting measured in parallel and normal (in brackets) directions with respect to the laser microgrooves.

| Rolling-off angle /° | water | diiodo-methane | humic acid | sodium alginate | bovine serum |
|---|---|---|---|---|---|
| *glass ref.* | 57 | 43 | 41 | 47 | 47 |
| *L -10(5)* | - | - | - | - | - |
| *F-L -10(5)* | 90 (90) | 49 (47) | 90 (72) | 90 (90) | 90 (90) |
| *L -25(5)* | - | - | - | - | - |
| *F- L -25(5)* | 90 (90) | 42 (20) | 30 (90) | 40 (90) | 20 (90) |
| *L-50(5)* | - | - | - | - | - |
| *F- L-50(5)* | 90 (90) | 62 (35) | 55 (90) | 62 (90) | 60 (90) |
| *L-100(5)* | - | - | - | - | - |
| *F- L-100(5)* | 90 (90) | 75 (30) | 45 (90) | 55 (90) | 60 (90) |
| *L-200(5)* | - | - | - | - | - |
| *F- L-200(5)* | 90 (90) | 15 (15) | 60 (90) | 60 (90) | 57 (90) |
| *L-300(5)* | - | - | - | - | - |
| *F- L-300(5)* | 90 (90) | 16 (90) | 44 (90) | 45 (90) | 50 (90) |
| *L-500(5)* | - | - | - | - | - |
| *F- L-500(5)* | 23 (90) | 39 (90) | 40 (90) | 40 (90) | 40 (90) |

**Supporting information S9.- *Rolling off (RoA) angles of different biological fluid simulant droplets on samples C and F-C.***

**Table S5**. Rolling-off angle values of 15 µl droplets of different liquids on samples *C* and *F-C* as function of the microgroove separation.

| Rolling-off angle /º | water | diiodo-methane | humic acid | sodium alginate | bovine serum |
|---|---|---|---|---|---|
| *Glass ref* | 57 | 43 | 41 | 47 | 47 |
| *C-10(5)* | - | - | - | - | - |
| *F-C-10(5)* | 90 | 90 | 87 | 90 | 90 |
| *C-25(5)* | - | - | - | - | - |
| *F- C-25(5)* | 90 | - | 90 | 90 | 90 |
| *C-50(5)* | - | - | - | - | - |
| *F- C-50(5)* | <2 | 70 | <2 | <2 | <2 |
| *C-100(5)* | - | - | - | - | - |
| *F- C-100(5)* | <2 | 79 | <2 | <2 | <2 |
| *C-200(5)* | - | - | - | - | - |
| *F- C-200(5)* | 90 | 60 | 90 | 90 | 90 |
| *C-300(5)* | - | - | - | - | - |
| *F- C-300(5)* | 90 | 30 | 90 | 90 | 90 |
| *C-500(5)* | - | - | - | - | - |
| *F-C-500(5)* | 90 | 39 | 90 | 90 | 90 |

**Supporting information S10.-** ***"Open air" demonstration of the anti-icing capacity of laser treated and molecular grafted samples.***

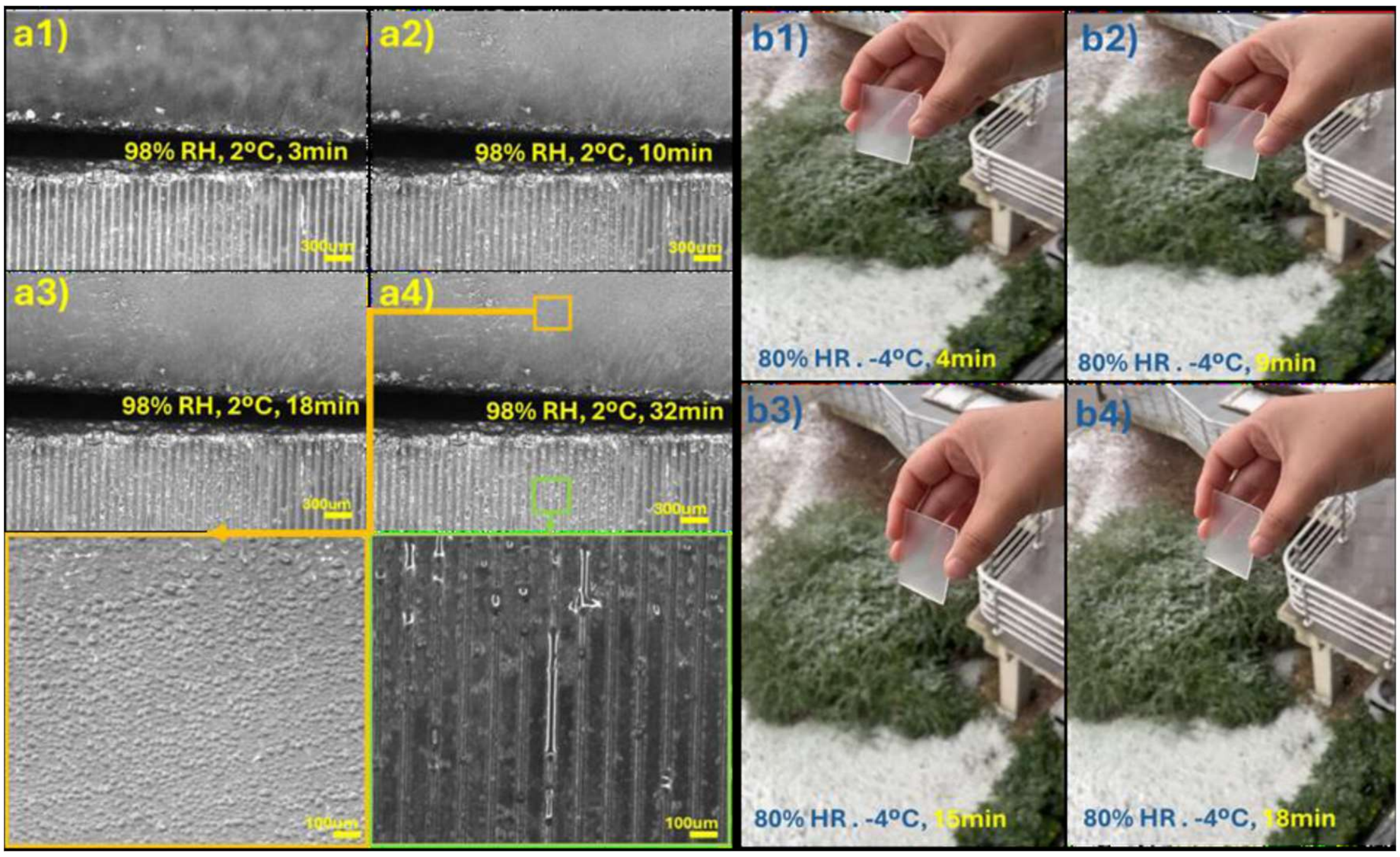

**Figure S7.** a) Sequence of images of a water condensation test in a controlled environmental chamber at 2°C and 48% RH and increasing times for a reference sample (top) and *F-L-100(5)* sample (bottom) including two zoomed regions of the last event in horizontal configuration; b) sequence of images of a real "open air" experiment with the *F-L-100(5)* sample at -4 °C for different snow exposure times.

**Video S1**. Real outdoor experiment with the *F-L-100(5)* sample at -4 °C.

**Supporting information S11.-** ***Water contact angles, freezing delay times and ice adhesion strengths of samples with and without infused Krytox.***

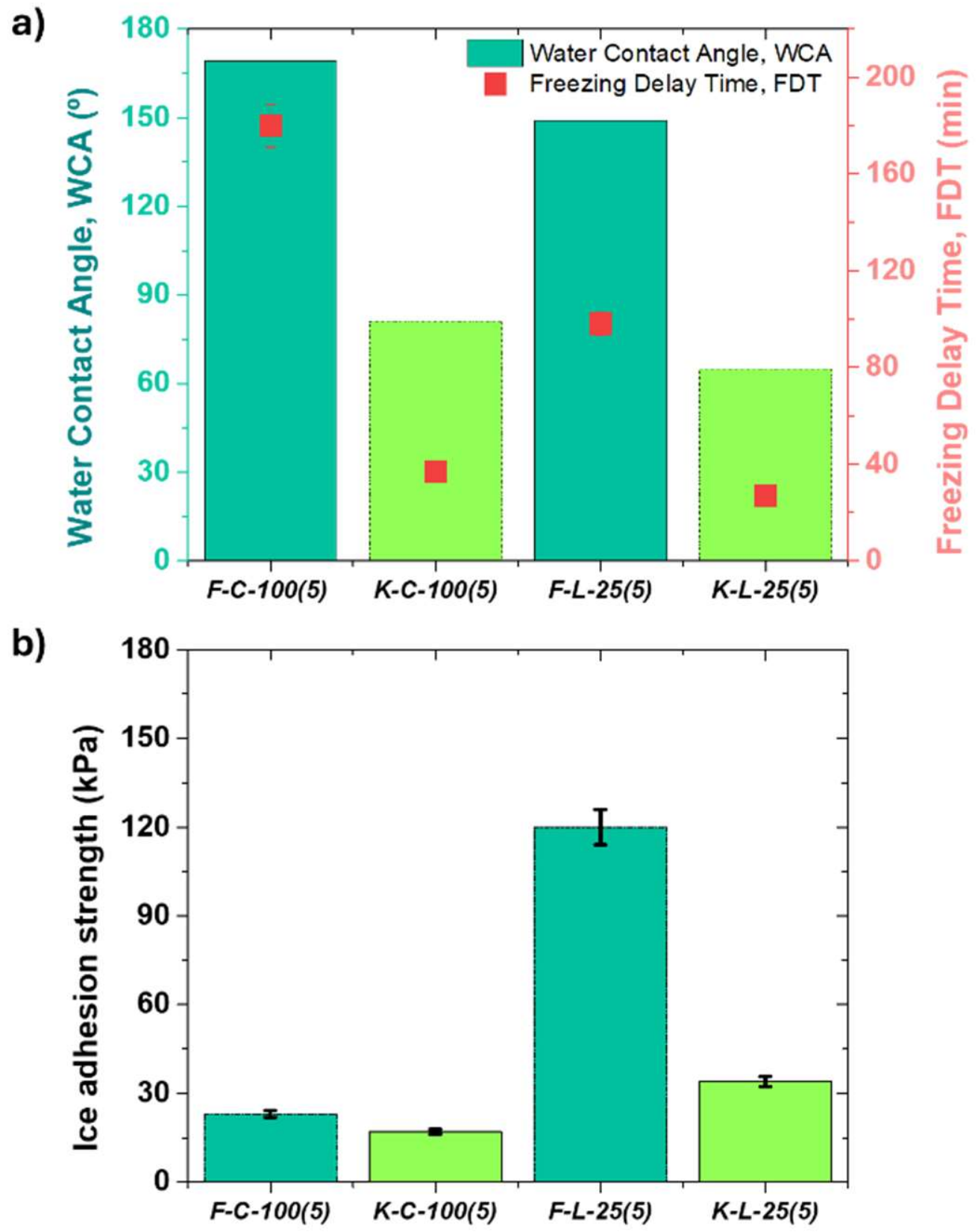

**Figure S8**. a) Water contact angles, and freezing delay times, and b) ice adhesion measurements, for SLIP surfaces after the infusion of Krytox lubricant. Comparison of samples *F-L-25(5), F-C-100(5)* with samples *K-L-25(5), K-C-100(5)*.